\documentclass{emulateapj}

\usepackage{ifthen}

\usepackage{verbatim}

\slugcomment{Last revision \today.}

\newcommand{\deltasig}{$\Delta \Sigma$}

\newcommand{\deltarho}{$\Delta \rho$}

\newcommand{\lstarlim}{$0.4 L_*$}
\newcommand{\lvir}{$L_{200}$}

\newcommand{\mvir}{$M_{200}$}
\newcommand{\nvir}{$N_{200}$}
\newcommand{\rvirgal}{$r_{200}^{gals}$}
\newcommand{\rvirmass}{$r_{200}^{mass}$}

\newcommand{\deltamtol}{$\Delta M/\Delta L$}
\newcommand{\deltam}{$\Delta M$}
\newcommand{\deltal}{$\Delta L$}

\newcommand{\deltamvir}{$\Delta M_{200}$}
\newcommand{\deltalvir}{$\Delta L_{200}$}

\newcommand{\mtolmax}{$(\Delta M/\Delta L)_{22\mathrm{Mpc}}$}
\newcommand{\mtolasym}{$(\Delta M/\Delta L)_{asym}$}
\newcommand{\mtolvir}{$(\Delta M/\Delta L)_{200}$}
\newcommand{\bmtol}{$b^2_{M/L}$}
\newcommand{\bmtolinv}{$b^{-2}_{M/L}$}

\newcommand{\maxbcg}{MaxBCG}

\newcommand{\zmean}{0.25}

\newcommand{\minrad}{$25h^{-1}$kpc}
\newcommand{\maxrad}{$30h^{-1}$Mpc}
\newcommand{\maxinvertrad}{$22h^{-1}$Mpc}

\newcommand{\meanmtolrmaxn}{$362 \pm 54 h$}
\newcommand{\meanmtolrmaxl}{$349 \pm 51 h$}
\newcommand{\meanmtolasymn}{$357 \pm 9 h$}
\newcommand{\meanmtolasyml}{$335 \pm 9 h$}

\newcommand{\omegarmax}{$0.20 \pm 0.03$}
\newcommand{\omegaasym}{$0.194 \pm 0.008$}

\newcommand{\lthresh}{$10^{9.5} h^{-2} L_\odot$}
\newcommand{\mthresh}{-19.08}

\newcommand{\ishift}{$^{0.25}i$}
\newcommand{\gmrshift}{$^{0.25}(g-r)$}

\newcommand{\lshift}{$L_{^{0.25}i}$}
\newcommand{\mshift}{$M_{^{0.25}i} - 5\log_{10}h$}

\newcommand{\ldens}{$(3.14 \pm 0.10) \times 10^8$ $h$ $L_\odot$Mpc$^{-3}$}

\newcommand{\nvirsampdef}{$12 \leq $\nvir$ \leq 17$}

\newcommand{\rhalf}{$r_{1/2}$}

\newcommand{\indexmtolvsm}{$0.33 \pm 0.02$}

\shortauthors{Sheldon et al.}
\shorttitle{SDSS Cluster M/L}

\begin{document}

\title{Cross-correlation Weak Lensing of SDSS Galaxy Clusters III: Mass-to-light Ratios}

\author{
Erin S. Sheldon,\altaffilmark{1}
David E. Johnston,\altaffilmark{2,3}
Morad Masjedi,\altaffilmark{1}
Timothy A. McKay,\altaffilmark{4,5,6}
Michael R. Blanton,\altaffilmark{1}
Ryan Scranton,\altaffilmark{7}
Risa H. Wechsler,\altaffilmark{8}
Benjamin P. Koester,\altaffilmark{9,10}
Sarah M. Hansen,\altaffilmark{9,10}
Joshua A. Frieman,\altaffilmark{9,10,11}
and James Annis\altaffilmark{11}
}

\altaffiltext{1}{Center for Cosmology and Particle Physics, Department of Physics, New York University, 4 Washington Place, New York, NY 10003.}
\altaffiltext{2}{Department of Astronomy, 105-24, California Institute of Technology, 1201 East California Boulevard, Pasadena, CA 91125.}
\altaffiltext{3}{Jet Propulsion Laboratory 4800 Oak Grove Drive, Pasadena ,CA, 91109}
\altaffiltext{4}{Department of Physics, University of Michigan, 500 East University, Ann Arbor, MI 48109-1120.}
\altaffiltext{5}{Department of Astronomy, University of Michigan, 500 Church St., Ann Arbor, MI 48109-1042}
\altaffiltext{6}{Michigan Center for Theoretical Physics, University of Michigan, 500 Church St., Ann Arbor, MI 48109-1042}
\altaffiltext{7}{Department of Physics and Astronomy, University of Pittsburgh, 3941 O'Hara Street, Pittsburgh, PA 15260.}
\altaffiltext{8}{Kavli Institute for Particle Astrophysics \& Cosmology, Physics Department, and Stanford Linear Accelerator Center, Stanford University, Stanford, CA 94305}
\altaffiltext{9}{Kavli Institute for Cosmological Physics, The University of Chicago, 5640 South Ellis Avenue Chicago, IL  60637}
\altaffiltext{10}{Department of Astronomy and Astrophysics, The University of Chicago, 5640 South Ellis Avenue, Chicago, IL 60637.}
\altaffiltext{11}{Fermi National Accelerator Laboratory, P.O. Box 500, Batavia, IL 60510.}

\begin{abstract}

We present measurements of the excess mass-to-light ratio measured around
\maxbcg\ galaxy clusters observed in the SDSS.  This red sequence cluster
sample includes objects from small groups with $M_{200} \sim 5\times10^{12}
h^{-1} M_{\odot}$ to clusters with $M_{200} \sim 10^{15} h^{-1} M_{\odot}$.
Using cross-correlation weak lensing, we measure the excess mass density
profile above the universal mean $\Delta \rho(r) = \rho(r) - \bar{\rho}$ for
clusters in bins of richness and optical luminosity. We also measure the excess
luminosity density $\Delta \ell(r) = \ell(r) - \bar{\ell}$ measured in the
$z=0.25~i$-band.  For both mass and light, we de-project the profiles to
produce 3D mass and light profiles over scales from $25 h^{-1}$ kpc to $22
h^{-1}$ Mpc.  From these profiles we calculate the cumulative excess mass
\deltam$(r)$ and excess light \deltal$(r)$ as a function of separation from the
BCG. On small scales, where $\rho(r) \gg \bar{\rho}$, the integrated
mass-to-light profile $($\deltamtol$)(r)$ may be interpreted as the cluster
mass-to-light ratio.  We find the \mtolvir, the mass-to-light ratio within
$r_{200}$, scales with cluster mass as a power law with index \indexmtolvsm.
On large scales, where $\rho(r) \sim \bar{\rho}$, the \deltamtol\ approaches an
asymptotic value independent of cluster richness.  For small groups, the mean
\mtolvir\ is much smaller than the asymptotic value, while for large clusters
\mtolvir\ is consistent with the asymptotic value.  This asymptotic value
should be proportional to the mean mass-to-light ratio of the universe $\langle
M/L \rangle$.  We find $\langle M/L \rangle~\textrm{\bmtolinv} =
\textrm{\meanmtolrmaxn}$ (statistical). There is additional uncertainty in the
overall calibration at the $\sim$10\% level. The parameter \bmtol\ is primarily
a function of the bias of the $L \lesssim L_*$ galaxies used as light tracers,
and should be of order unity. Multiplying by the luminosity density in the same
bandpass we find $\Omega_m \textrm{\bmtolinv} = \textrm{\omegarmax}$,
independent of the Hubble parameter.

\end{abstract}


\section{Introduction} \label{sec:intro}

Comparison of luminous to total mass through the mass-to-light ratio (M/L)
dates back at least to \citet{Kapteyn22}, who found values of M/L $\sim$ 2
M$_\odot$/L$_\odot$ in the solar neighborhood. Similar values were found by
Hubble for the disk of Andromeda in 1929 \citep{Hubble29}. Extension of this
approach to larger scales was first attempted by Zwicky in 1933 for the Coma
cluster \citep{Zwicky33,Zwicky37}.  Under reasonable assumptions about the
cluster's dynamical state, Zwicky estimated the mass of the cluster.  The
inferred average M/L for cluster galaxies was around 500 M$_\odot$/L$_\odot$,
implying a surprising dominance of dark over luminous matter. 

Decades of observation confirm Zwicky's conclusion that dark matter dominates
on large scales. The simplest model capable of accommodating current
observations is the Cold Dark Matter (CDM) model. It is assumed that
approximately 80\% of the cosmic mass density is in the form of some
dynamically cold, collisionless species of subatomic particle. Monte Carlo
predictions for the evolution of structure in such a CDM universe have reached
a high level of sophistication.  Given a set of cosmological parameters and
initial conditions, modern simulations are accurate and predictive within well
understood limitations imposed by resolution and simulation volume \citep*[see
e.g.  the comparison paper of ][and references therein]{HeitmannCompare07}.
However, accurate simulations of the baryonic components of clusters remain
beyond both current theoretical understanding and simulation technology.  In
the absence of direct theoretical prediction, observations of the connection
between observable light and computable dark matter remain important. The
simplest connection, still relevant 90 years after Kapteyn, is the
mass-to-light ratio.

Observations of both mass and light have advanced considerably since Zwicky's
time. Dynamics of cluster galaxies still provide important constraints on
cluster masses. These have confirmed Zwicky's essential conclusions with
improved accuracy over a broader range of environments
\citep*[e.g.][]{CarlbergM2l97}.  There are also new analysis techniques such as
the ``caustics'' method \citep{RinesCairnsM2l04} and velocity stacking methods
\citep*[e.g][]{BeckerVelocities07} which confirm and extend the standard
analyses. 

The X--ray emission from the hot intra-cluster plasma has become a favored
proxy for cluster mass. Under the assumption of hydrostatic equilibrium, masses
and mass-to-light ratios can be calculated from X-ray temperatures and, less
directly, luminosities. These X-ray analyses confirm the dynamical measurements
of high mass-to-light ratios \citep*[e.g.][]{LinMohrStanford04}.

Weak gravitational lensing, an effect which Zwicky thought promising
\citep{Zwicky37}, is now an established technique for studying the mass
distributions in clusters of galaxies.  The effect, deflection of light from
background objects as it passes foreground clusters, is generally less precise
on a cluster-by-cluster basis than X--ray or velocity measurements. But it is
complementary in that it is independent of the dynamical state of the system,
is linear in the density, and can be measured at a large range of separations
from the cluster center.

Weak lensing studies have traditionally focused on measuring the mass of
individual clusters\footnote{A few examples:
\citet{Fahlman94,TysonFischer95,Luppino97,Fischer97,Hoekstra98,
Joffre00,Clowe00,Dahle02,Wittman03,Umetsu05,CloweDMProof06}.}. However,
averaging the lensing signal over many clusters is more robust and easier to
interpret \citep*[see the discussions in
][]{JohnstonInvert07,SheldonLensing07,JohnstonLensing07}.  With this method,
the full M/L profile can be measured from small scales to well beyond the bound
regions of the clusters.  Early stacked M/L results for group-sized objects
\citep[e.g.]{Hoekstra01c,Parker05} showed the promise of this approach.  The
pilot study of \citet{Sheldon01} demonstrated the great potential of the Sloan
Digital Sky Survey \citep[SDSS;][]{York00} for measuring cluster masses.

In this work we measure the mean mass-to-light ratios for SDSS groups and
clusters drawn from the \maxbcg\ cluster sample
\citep{KoesterAlgorithm07,KoesterCatalog07}. We compare the ensemble mass
estimates from the first two papers in this series
\citep{SheldonLensing07,JohnstonLensing07} with ensemble measurements of the
total light.  We measure the mean mass using lensing, and the mean luminosity
by correlating the clusters with the surrounding galaxies.  Each measurement
spans a range of separations from 25$h^{-1}$kpc to \maxinvertrad, extending the
volume over which the M/L is measured well beyond the virial radii of the
clusters.

Because lensing is not sensitive to uniform mass distributions, aka ``mass
sheets'', we measure the mean mass density of a lens sample above the mean
density of the universe \deltarho=$\rho(r) - \bar{\rho}$ (actually we measure a
2D projection of the 3D \deltarho\ which we will de-project).  In other words,
we measure the cluster-mass cross correlation function times the mean density
of the universe $\Delta \rho = \bar{\rho} \xi_{cm}$.  On small scales this is a
measure of the mean density profile, but on large scales, where the density
approaches the background, it can only be interpreted in terms of the
correlation function.

We measure the light using a stacking technique directly comparable to the
lensing measurements.  We include light from all galaxies surrounding clusters
with luminosity above a threshold, and then subtract the measurements around
random points in order to remove the uniform background.  This means we measure
the luminosity density above the mean $\Delta \ell=\ell(r) - \bar{\ell}$, so it
is a correlation function just like the mass measurement .  The excess mass
within radius $r$ divided by the excess luminosity within radius $r$ is the
ratio of integrals of correlation functions:
\begin{equation} \label{eq:m2ldef}
    \frac{\Delta M}{\Delta L}\left(<r\right) = 
    \frac{\int_0^r \mathrm{d}r\left[~\rho(r) - \bar{\rho}~\right]~r^2}
    { \int_0^r \mathrm{d}r\left[~\ell(r) - \bar{\ell}~\right]~r^2} = 
\frac{ \int_0^r \mathrm{d}r  ~\bar{\rho}~\xi_{cm}~r^2  }{
\int_0^r \mathrm{d}r ~\bar{\ell}~\xi_{c\ell} ~r^2 }.  
\end{equation} 

On the scale of virialized halos, defined where the mean density is a few
hundred times the mean, equation \ref{eq:m2ldef} can be interpreted as the mean
cluster M/L in a straightforward way. On very large scales, as the density
approaches the mean, equation \ref{eq:m2ldef} becomes proportional to the mean
mass-to-light ratio of the universe.  The proportionality constant is  related
to the particulars of the correlation function measurements.  For example, the
light and mass may be clustered differently , and this difference may depend on
the properties of the galaxies chosen as tracers of the light.  Thus we expect
this measurement to depend on the ``bias'' of the light tracers relative to the
mass.  This bias depends primarily on the mass of the halos hosting these
tracers. The bias also depends on the variance of the mass density field
($\sigma_8$) as the larger the variance in that field the smaller the bias at
fixed mass \citep{Kaiser84}.  


Full theoretical models of this measurement on all scales will require
substantial effort.  For this paper we will parametrize our ignorance at large
scales in terms of the bias.  We will write the asymptotic M/L as 
\begin{eqnarray} \label{eq:m2lbias_test}
    \textrm{\mtolasym} & = & \langle M/L \rangle~ \textrm{\bmtolinv} \nonumber \\
    \textrm{\bmtolinv} & = & \frac{b_{cm}}{b_{c\ell}} \frac{1}{b^2_{\ell m}}.
\end{eqnarray}
Here $\langle M/L \rangle$ is the mean mass-to-light ratio of the universe, but
it is multiplied by a bias factor \bmtolinv.  This factor depends on the bias
of the galaxy tracers relative to the mass $b_{\ell m}$, which should be
of order unity for tracers near $L_*$ since these galaxies are expected to be
only slightly anti-biased \citep{ShethTormen99,SeljakWarren04}.  This term also
depends on the ratio of the bias of clusters relative to mass and light
$b_{cm}/b_{c\ell}$, which should also be near unity for the same reasons.  The
cluster bias is likely to cancel from this equation because the cluster terms
appear in a ratio $b_{cm}/b_{c\ell}$.

On large enough scales, \deltamtol\ should approach a constant value
independent of the halo mass as long as the bias of the light tracers, and thus
\bmtol, is equivalent in all cases. This appears to be the case, as we will
demonstrate.

We assume a Friedman-Robertson-Walker cosmology with $\Omega_M$ = 0.27,
$\Omega_{\Lambda}$ = 0.73, and H$_0$ = 100 $h$ km/s/Mpc and distances are
measured in physical, or proper units, rather than comoving units.  

\newpage
\section{Methods} \label{sec:methods}

\subsection{Lensing Methods} \label{sec:lensmethods}

We will briefly describe the lensing measurements as these were described in
detail in \citet{SheldonLensing07} and \citet{JohnstonLensing07}.  We measure
the tangential shear induced in background galaxies by a set of foreground
clusters and convert that shear to a redshift independent density contrast
\deltasig:
\begin{equation} \label{eq:gammat}
\gamma_T (R) \times \Sigma_{crit} = \bar{\Sigma}(< R)
-\bar{\Sigma}(R) \equiv \Delta \Sigma ~,
\end{equation}
where $\bar{\Sigma}$ is the projected surface mass density at radius $R$ and
$\bar{\Sigma}(<R)$ is the mean projected density within radius $R$.  In using
this equation, we assume the shear is weak.  This is not always the case for
the largest clusters on the smallest scales, as was discussed in
\citet{JohnstonLensing07}.

We average \deltasig\ for an ensemble of clusters in radial bins from \minrad\
to \maxrad, using the brightest cluster galaxy (BCG) as the center for all
measurements (see \S \ref{sec:cluster_sample} for a description of the cluster
selection). The \deltasig\ is linear in the density and so averaging \deltasig\
averages the density directly (again assuming weak shear).  

A number of corrections were made to the profile, as discussed in detail in
\citep{SheldonLensing07}. Using random points, corrections were applied for
contamination of the background source sample with cluster members that are not
sheared; this correction is large but well-determined on small scales, and
negligible on large scales.   We also corrected for residual additive biases in
the shear.  These result from imperfect correction for biases in the galaxy
shapes caused by PSF anisotropy.  We use random points for this correction as
well, as the additive biases will appear in random points as well as around
clusters. This correction is negligible on small scales but significant on
large scales.

There are remaining uncertainties in the overall calibration of the mean
\deltasig.  These come primarily from uncertainties in photometric redshift
determinations for the background source galaxies and correction for blurring
of the galaxy shapes by the PSF.  Simulations of the PSF correction suggest it
is good to a few percent under simplified circumstances \citep{STEP2}. The
photometric redshift calibrations are probably less well constrained.
Comparisons with studies using Luminous Red Galaxies as sources, for which the
redshift is better determined, suggest the calibrations are good to about 10\%
\citep{MandelbaumSystematics05}, although that study used a different photoz
algorithm that this work.  Simulations of the algorithm used in this study also
suggest $\sim$10\% errors given the size of the training set \citep{Lima08}.

In \S \ref{sec:mass_profiles} we will present inversions of \deltasig\ to the
integrated excess mass.  These mass profiles are model-independent, but as with
the light measurements presented in \S \ref{sec:clmethods}, only the excess
above the mean can be measured with lensing.

\subsection{Cluster-light Correlations} \label{sec:clmethods}

We used the method of \citet{Masjedi06a} to estimate the mean number density
and luminosity density of galaxies around clusters.  This method is essentially
a correlation function with units of density: corrections are made for random
pairs along the line of sight as well as pairs missed due to edges and holes.
We defined two samples: The primary sample denoted $p$ and the secondary sample
denoted $s$.   In our case, the primaries were galaxy clusters with redshift
estimates and the secondaries were the imaging sample with no redshift
information, but in what follows we will use a more general notation.  For
example, the counts of real data secondaries around real data primaries is
denoted $D_p D_s$, while the counts of real secondaries around random primaries
is $R_p D_s$.  If instead of counting ``1'' for each object, we count some
other quantity such as the luminosity of the secondary, we say we have weighted
the measurement by luminosity.

The mean luminosity density of secondaries around primaries is 
\begin{equation} \label{eq:estimator}
\bar{\ell}~ w = \frac{D_p D_s}{D_p R_s} - \frac{R_p D_s}{R_p R_s},
\end{equation}
where $\bar{\ell}$ is the mean luminosity density of the secondary sample,
averaged over the redshift distribution of the primaries, and $w$ is the
projected correlation function.  This is the estimator from \citet{Masjedi06a}
where the weight of each primary-secondary pair is the luminosity of the
secondary.  We have written the measurement as $\bar{\ell} w$ to illustrate
that the estimator gives the mean density of the secondaries times the
projected correlation function $w$. Only the excess luminosity density with
respect to the mean can be measured.  Using a weight of unity gives the number
density.  

The first term in equation \ref{eq:estimator} estimates the total luminosity
density around the clusters including everything from the secondary imaging
sample projected along the line of sight, and the second term corrects for the
random pairs along the line of sight.  Note, the same secondary may be counted
around multiple primaries (or random primaries).

The numerator of the first term in equation \ref{eq:estimator}, $D_p D_s$, is 
calculated as:
\begin{equation}
D_p D_s = \frac{\sum_{p,s}{L_s}}{N_p} 
  = \langle L_{pair} \rangle + \langle L^R_{pair} \rangle 
  = f A~\bar{\ell}~(w + 1),
\end{equation}
where the sum is over all pairs of primaries and secondaries, weighted by the
luminosity of the secondary.  The secondary luminosity is calculated by
K-correcting each secondary galaxy's flux assuming it is at the same redshift
as the primary (see \S \ref{sec:kcorr} for details of the K-corrections).   
The total luminosity is
the sum over correlated pairs ($L_{pair}$) as well as random pairs along the
line of sight ($L^R_{pair}$). By the definition of $w$, this is the total
luminosity per primary times $w+1$.  This can be rewritten in terms of the
luminosity density of the secondaries $\bar{\ell}$ times the area probed $A$.
Some fraction of the area searched around the lenses is empty of secondary
galaxies due to survey edges and holes. The factor $f$ represents the mean
fraction of area around each primary actually covered by the secondary catalog.
It is a function of pair separation, with a mean value close to 1 on small
scales but then dropping rapidly at large scales.

Again, a single secondary may be counted around multiple primaries and
K-corrected to different redshifts. Statistically they will only contribute
when paired with a physically associated object due to the background
subtraction described below. In fact most of the calculations involved in this
measurement are for pairs that are not physically associated, which is part of
the reason this is computationally difficult.

The denominator of the first term in Equation \ref{eq:estimator} calculates the
factor $f A$, the actual area probed around the primaries.  This term in the
denominator corrects for the effects of edges and holes. Also, because it has
units of area, we recover the density rather than just the correlation
function.
\begin{equation}
D_p R_s = \frac{ N^{DR}_{pair} }{ \sum_{p} \left( \frac{d\Omega}{dA} \right)_p \frac{dN}{d\Omega}  }
  = f A.
\end{equation}
The numerator is the pair counts between primaries and random secondaries, and
the denominator is the expected density of pairs averaged over the redshift
distribution of the primaries, times the number of primaries.  The ratio is the
actual mean area used around each primary $f A$.

The second term in equation \ref{eq:estimator} accounts for the
random pairs along the line of sight.  The numerator and denominator
are calculated the same way as the first term in equation \ref{eq:estimator},
but with fake primaries distributed randomly over the survey geometry.
The redshifts are chosen such that the distribution of redshifts
smoothed in bins of $\Delta z = 0.01$ match that of the clusters.
\begin{eqnarray}
R_p D_s = \frac{\sum_{pr,s}{L_s}}{N_p} 
  = \langle L^R_{pair} \rangle 
  = f^R A^R~\bar{\ell} \label{eq:rd} \\
R_p R_s = \frac{ N^{RR}_{pair} }{ \sum_{pr} \left( \frac{d\Omega}{dA} \right)_{pr} \frac{dN}{d\Omega}  }
  = f^R A^R.
\end{eqnarray}
The ratio of these two terms, $R_p D_s / R_p R_s$, calculates the mean 
density of the secondaries after correcting for the survey geometry.

The density measured with this technique could be tabulated in various ways,
typically as a function of projected radial separation $R$.  We tabulate in a
cube which represents bins of separation $R$, luminosity $L$, and color $g-r$.
This facilitates the study of the radial dependence of the luminosity function
and the color-density relation.

These profiles can be inverted to the three-dimensional excess density and
integrated to get the total excess light.  We will present this formalism in
section \ref{sec:lum_profiles}.

\section{Data}

The data used for lensing was described in detail in paper I
\citep{SheldonLensing07} and the cluster sample is described in
\citet{KoesterAlgorithm07,KoesterCatalog07} with the modifications detailed in
\citet{SheldonLensing07}.  We will briefly describe relevant details of the
cluster sample and give a full description of the galaxies used as light
tracers in the cluster-light cross-correlation measurements.  All the primary
data in this study come from the Sloan Digital Sky Survey
\citep[SDSS;][]{York00} data release 4 \citep{dr4}.

\subsection{Cluster Sample} \label{sec:cluster_sample}

Full details of the cluster finder and catalog can be found in
\citet{KoesterAlgorithm07,KoesterCatalog07}. The cluster finder is a
red-sequence method, limited to the redshift range 0.1-0.3.  The basic galaxy
count \nvir\ is the number of galaxies on the red sequence with rest-frame
i-band luminosity $L >$\lstarlim\ within \rvirgal.  The $i$-band $L_*$ used in
the cluster finder is the $z=0.1$ value from \citet{BlantonLum03},
corresponding to $M_* -5$log$_{10}h = -20.82 \pm 0.02$.  The radius \rvirgal\
is determined from the size-richness relation presented in \citet{Hansen05}.
Note, this relation gives roughly a factor of 2 larger radius than the
\rvirmass\ determined from the mass in Paper II \citep{JohnstonLensing07} (see
\S \ref{sec:intro} for details about the various richness measures).  The
published catalog contains clusters with \nvir\ $\geq 10$, and we augment this
catalog with \nvir\ $\geq 3$ objects.  The cluster photometric redshifts are
accurate to 0.004 over our redshift range with a scatter of $\Delta z \sim
0.01$ for \nvir\ $\geq 10$; the scatter degrades to $\Delta z \sim 0.02$ for
\nvir\ $\geq 3$, with the same accuracy.

\subsection{Galaxy Sample} \label{sec:galaxy_sample}

For the cluster-light cross-correlations we separated galaxies from stars using
the Bayesian techniques developed in \citet{Scranton02}. The primary source of
confusion in star-galaxy separation at faint magnitudes is shot noise.  Stars
scatter out of the stellar locus and galaxies scatter into the stellar locus.
This technique uses knowledge of the true size distribution of stars and
galaxies as a function of apparent magnitude to assign each object a
probability of being a galaxy. 

We characterized the distribution of star and galaxy sizes as a function of
magnitude and seeing using the deeper southern SDSS stripe. The SDSS southern
stripe has a been repeatedly scanned. We chose regions of sky which were
scanned at least 20 times and chose the 20 best-seeing observations for each
object.  We then simply added the flux at the catalog level to increase the
S/N.  Thus, the selection is close to that of single scans but with a better
measurement; S/N at a given magnitude is higher, and the distribution of
measured sizes is closer to the truth.  For a range of seeing values, we then
calculated the probability that an object in a single-epoch image with a given
magnitude and size is truly a galaxy, and applied this to all objects in the
survey.  The resulting distribution is highly peaked at probability 0 and 1,
such that our chosen probability cut at $p >$ 0.8 results in a sample 
$\gtrsim$99\% pure within our magnitude limit.

We chose to K-correct to band-passes shifted to the mean cluster redshift 0.25
rather than redshift 0 to minimize the corrections.  We will refer to all such
magnitudes with a superscript, e.g.  \ishift. For a full discussion of the
band-shifting process see \citet{BlantonKcorr03}.

We chose a volume and magnitude limited galaxy sample for \lshift\ $ >
$\lthresh\ (or \mshift\ $ < $\mthresh).  and $z < 0.3$.  This corresponds to an
apparent magnitude limit of $i < 21.3$, and color limits of $g-r < 2$ and $r-i
< 1$.  All magnitudes are SDSS model magnitudes.

\subsection{Survey Geometry} \label{sec:window}

We characterized the survey geometry using the SDSSPix code
\footnote{http://lahmu.phyast.pitt.edu/$\sim$scranton/SDSSPix/}. This code
represents the survey using nearly equal area pixels, including edges and holes
from missing fields and ``bad'' areas near bright stars.  We removed areas with
extinction greater than 0.2 magnitudes in the dust maps of \citet{Schlegel98}.
This window function was used in the cluster finding and in defining the galaxy
catalog for the cross correlations.  By including objects only from within the
window, and generating random catalogs in the same regions, we controlled and
corrected for edges and holes in the observed counts as described in \S
\ref{sec:methods}. 

\subsection{K-corrections} \label{sec:kcorr}

We calculated K-corrections using the template code \texttt{kcorrect} from
\citet{BlantonKcorr03}.  This code is accurate but too slow to calculate the
K-corrections for the billions of pairs found in the cross-correlations. To
save time we computed the K-correction on a grid of colors in advance.  We took
galaxies from the SDSS Main sample as representative of all galaxy types.  We
then computed their K-corrections on a grid of redshifts between 0 and 0.3, the
largest redshift considered for clusters in this study.  The mean K-correction
in a $21\times21\times80$ grid of observed $g-r, r-i$, and $z$ was saved.  We
interpolated this cube when calculating the K-correction for each neighboring
galaxy. This interpolation makes the calculation computationally feasible for
this study, but is still the bottleneck.

\section{Cluster Galaxy Population Measurements}

\subsection{Radial, Color and Luminosity Binning} \label{sec:galbin}

Using the estimator presented in \S \ref{sec:clmethods}, we measured the number
and luminosity density of galaxies around the clusters as a function of radius
$R$ from the BCG, color \gmrshift, and luminosity \lshift.  The bins in radius,
color, and luminosity form a data cube with 18 bins in radius, 20 bins in color
and 20 bins in luminosity.  The range in each variable is $0.02<R<11.5 h^{-1}$
Mpc, $0<$\gmrshift$<2$ and $9.5<$log$_{10}($\lshift$/L_{\odot})<11.7$.   The
initial cube before background correction was kept separately for each cluster
to allow flexibility when later estimating the average profile. Two versions of
this cube were kept, one with number counts and another with the total
luminosity.  In other words, in the first case we counted each galaxy and in
the other we counted the luminosity.  Another set of measurements was also
performed to $30 h^{-1}$ Mpc, but binning only in radial separation to save
resources.   More detailed analysis of the full cube will presented in
\citet{Hansen07}; here we will present what is needed for the particulars of
the M/L measurement.

\subsection{Random Catalogs} \label{sec:randoms}

We generated random catalogs uniformly over the survey area using the window
function described in \S \ref{sec:window}. We chose the redshift distribution
to be that of constant density in comoving volumes over the redshift range of
the clusters. We matched subsets of these redshifts to the redshift
distribution of each cluster subsample as described in \S \ref{sec:histmatch}. 

We performed the same galaxy counting as described in \S \ref{sec:galbin} for a
set of 15 million random points in order to correct for the random background.
These are used in the RD term from the estimator described in \S
\ref{sec:methods} (Equations \ref{eq:estimator} and \ref{eq:rd}). We also ran
sets of 15 million random points for the DR and RR terms.

\subsection{Histogram Matching} \label{sec:histmatch}

\looseness+1 The redshifts used for random points must statistically match that
of each cluster sample in order for the background subtraction to be accurate.
The random primaries described in \S \ref{sec:randoms} were generated with
constant comoving density.  We drew random redshifts from this sample such that
the distribution matched that of the cluster sample when binned with 
$\Delta z=0.01$.

\subsection{Cluster Richness Binning} \label{sec:clusterbin}

Because the measurements before background correction were saved for each
cluster separately, the clusters could be binned at a later time to produce
mean density profiles.  For this work we binned the clusters into 12 bins of
\nvir\ and 16 bins of $i$-band cluster luminosity \lvir, where the luminosity
is that of the \nvir\ red galaxies counted within \rvirgal\ as described in \S
\ref{sec:cluster_sample}.  These bins were chosen to correspond with the
binning presented in the lensing analysis of \citet{SheldonLensing07}.  Some
statistics of this sample are shown in Tables \ref{tab:m2lngals} and
\ref{tab:m2llum}. 

\begin{deluxetable*}{ccccccccc}
\tabletypesize{\scriptsize}
\tablecaption{\deltamtol\ Statistics for \nvir\ Bins \label{tab:m2lngals}}
\tablewidth{0pt}
\tablehead{
    \colhead{$\langle$\nvir $\rangle$}       &
    \colhead{$\Delta L_{200}$}         &
    \colhead{$\Delta M_{200}$}  &
    \colhead{$\left(\frac{\Delta M}{\Delta L}\right)_{200}$}  &
    \colhead{$\left(\frac{\Delta M}{\Delta L}\right)_{rmax}$}  &
    \colhead{$\langle L^{gal}\rangle $} &
    \colhead{\rhalf} &
    \colhead{$\alpha$} &
    \colhead{$\left(\frac{\Delta M}{\Delta L}\right)_{asym}$}  
}
\
\startdata
3 & $  3.84 \pm   0.02$ & $  3.99 \pm   0.16$ & $101 \pm 10$ & $275 \pm 313$ & $0.98 \pm 0.10$ & $ 0.25 \pm  0.07$ & $ 1.34 \pm  0.11$ & $226 \pm 33$  \\
4 & $  4.86 \pm   0.04$ & $  5.78 \pm   0.26$ & $108 \pm 14$ & $347 \pm 229$ & $1.02 \pm 0.10$ & $ 0.44 \pm  0.15$ & $ 1.05 \pm  0.08$ & $306 \pm 42$  \\
4 & $  5.98 \pm   0.07$ & $  8.84 \pm   0.44$ & $166 \pm 18$ & $455 \pm 234$ & $1.02 \pm 0.11$ & $ 0.47 \pm  0.16$ & $ 1.06 \pm  0.08$ & $357 \pm 44$  \\
5 & $  7.22 \pm   0.09$ & $ 12.88 \pm   0.64$ & $170 \pm 22$ & $341 \pm 183$ & $0.93 \pm 0.12$ & $ 0.24 \pm  0.05$ & $ 1.46 \pm  0.13$ & $285 \pm 28$  \\
7 & $  7.99 \pm   0.12$ & $ 12.54 \pm   0.78$ & $122 \pm 27$ & $540 \pm 231$ & $1.10 \pm 0.11$ & $ 0.30 \pm  0.31$ & $ 1.02 \pm  0.14$ & $271 \pm 59$  \\
8 & $  9.16 \pm   0.15$ & $ 16.58 \pm   1.20$ & $179 \pm 30$ & $353 \pm 157$ & $1.10 \pm 0.11$ & $ 0.44 \pm  0.12$ & $ 1.16 \pm  0.11$ & $378 \pm 41$  \\
9 & $ 11.51 \pm   0.14$ & $ 26.88 \pm   1.03$ & $229 \pm 22$ & $354 \pm 140$ & $1.04 \pm 0.10$ & $ 0.30 \pm  0.05$ & $ 1.27 \pm  0.09$ & $342 \pm 24$  \\
13 & $ 16.23 \pm   0.22$ & $ 42.67 \pm   1.39$ & $257 \pm 21$ & $327 \pm 133$ & $1.08 \pm 0.09$ & $ 0.28 \pm  0.05$ & $ 1.12 \pm  0.08$ & $355 \pm 24$  \\
20 & $ 23.74 \pm   0.38$ & $ 70.18 \pm   2.54$ & $319 \pm 27$ & $321 \pm 145$ & $1.24 \pm 0.11$ & $ 0.19 \pm  0.03$ & $ 1.37 \pm  0.11$ & $349 \pm 21$  \\
31 & $ 36.88 \pm   0.80$ & $131.66 \pm   5.02$ & $333 \pm 32$ & $300 \pm 175$ & $1.27 \pm 0.13$ & $ 0.20 \pm  0.04$ & $ 1.35 \pm  0.16$ & $394 \pm 26$  \\
50 & $ 53.82 \pm   1.95$ & $203.71 \pm   9.57$ & $379 \pm 43$ & $525 \pm 307$ & $0.94 \pm 0.24$ & $ 0.15 \pm  0.02$ & $ 2.19 \pm  0.29$ & $404 \pm 21$  \\
92 & $104.33 \pm   5.16$ & $516.19 \pm  38.78$ & $489 \pm 75$ & $547 \pm 332$ & $0.93 \pm 0.53$ & $ 0.11 \pm  0.01$ & $ 3.91 \pm  0.65$ & $485 \pm 33$ 
\enddata

\tablecomments{Mass-to-light ratio statistics for clusters binned by richness
\nvir.  \deltal\ is the excess light over the mean luminosity density of the
universe, and \deltam\ is the excess mass over the mean mass density.
Subscripts $200$ and $rmax$ indicate quantities within $r_{200}$ and the
maximum radius \maxinvertrad, respectively.  The subscript $asym$ refers to an
asymptotic value from fitting the model described in the text.  Other
parameters of this fit are $r_{1/2}$ and $\alpha$, the half-light radius and
power law index.  The model is not a physical model, so these uncertainties
should be considered lower limits.  $L^{gal}$ is the mean luminosity of the
neighboring galaxies used to calculate \deltal. No attempt was made to account
for possible offsets between the BCGs used as centers and the halo mass peak;
this would affect the \deltamvir\ but should have little effect on \mtolvir,
and no effect on \mtolmax.  Masses are in units of $h^{-1} M_\odot$,
luminosities in units of $h^{-2} 10^{10} L_{\odot}$ and mass-to-light ratios in
units of $h M_\odot/L_\odot$.  \rhalf\ is in units of $h^{-1} $Mpc.  The mean
richness is shown but the ranges can be found in \citet{SheldonLensing07}}  

\end{deluxetable*}

\begin{deluxetable*}{ccccccccc}
\tabletypesize{\scriptsize}
\tablecaption{\deltamtol\ Statistics for \lvir\ Bins \label{tab:m2llum}}
\tablewidth{0pt}
\tablehead{
    \colhead{$\langle$\lvir $\rangle$}       &
    \colhead{$\Delta L_{200}$}         &
    \colhead{$\Delta M_{200}$}  &
    \colhead{$\left(\frac{\Delta M}{\Delta L}\right)_{200}$}  &
    \colhead{$\left(\frac{\Delta M}{\Delta L}\right)_{rmax}$}  &
    \colhead{$\langle L^{gal} \rangle $} &
    \colhead{\rhalf} &
    \colhead{$\alpha$} &
    \colhead{$\left(\frac{\Delta M}{\Delta L}\right)_{asym}$}  
}
\
\startdata
  5.6 & $  3.75 \pm   0.04$ & $  4.51 \pm   0.30$ & $114 \pm 20$ & $363 \pm 263$ & $1.04 \pm 0.11$ & $ 0.70 \pm  0.50$ & $ 0.94 \pm  0.08$ & $465 \pm 96$  \\
  7.0 & $  4.63 \pm   0.04$ & $  5.32 \pm   0.32$ & $113 \pm 18$ & $260 \pm 238$ & $1.03 \pm 0.11$ & $ 0.39 \pm  0.19$ & $ 1.15 \pm  0.11$ & $300 \pm 56$  \\
  8.7 & $  5.98 \pm   0.06$ & $  7.64 \pm   0.40$ & $116 \pm 18$ & $458 \pm 213$ & $1.01 \pm 0.11$ & $ 0.18 \pm  0.05$ & $ 1.44 \pm  0.15$ & $200 \pm 25$  \\
 10.8 & $  7.62 \pm   0.07$ & $ 11.80 \pm   0.53$ & $169 \pm 18$ & $283 \pm 146$ & $1.11 \pm 0.12$ & $ 0.42 \pm  0.12$ & $ 1.09 \pm  0.08$ & $315 \pm 36$  \\
 13.5 & $  9.48 \pm   0.10$ & $ 15.68 \pm   0.70$ & $129 \pm 19$ & $405 \pm 187$ & $1.05 \pm 0.11$ & $ 0.52 \pm  0.13$ & $ 1.06 \pm  0.06$ & $357 \pm 38$  \\
 16.9 & $ 11.82 \pm   0.13$ & $ 23.68 \pm   0.94$ & $193 \pm 19$ & $374 \pm 173$ & $1.11 \pm 0.11$ & $ 0.52 \pm  0.15$ & $ 0.98 \pm  0.07$ & $399 \pm 42$  \\
 21.1 & $ 14.12 \pm   0.19$ & $ 29.34 \pm   1.26$ & $214 \pm 23$ & $526 \pm 184$ & $1.16 \pm 0.11$ & $ 0.33 \pm  0.12$ & $ 1.08 \pm  0.10$ & $319 \pm 36$  \\
 26.3 & $ 17.77 \pm   0.27$ & $ 41.87 \pm   1.76$ & $215 \pm 25$ & $332 \pm 137$ & $1.15 \pm 0.11$ & $ 0.33 \pm  0.09$ & $ 1.13 \pm  0.10$ & $357 \pm 35$  \\
 32.9 & $ 21.75 \pm   0.35$ & $ 54.22 \pm   2.46$ & $265 \pm 27$ & $408 \pm 223$ & $1.02 \pm 0.12$ & $ 0.21 \pm  0.05$ & $ 1.33 \pm  0.13$ & $329 \pm 28$  \\
 40.9 & $ 26.39 \pm   0.56$ & $ 72.34 \pm   3.31$ & $279 \pm 32$ & $315 \pm 159$ & $1.30 \pm 0.15$ & $ 0.22 \pm  0.05$ & $ 1.28 \pm  0.13$ & $343 \pm 27$  \\
 51.2 & $ 33.96 \pm   0.84$ & $105.56 \pm   5.22$ & $284 \pm 39$ & $237 \pm 218$ & $1.24 \pm 0.16$ & $ 0.24 \pm  0.09$ & $ 1.15 \pm  0.15$ & $368 \pm 41$  \\
 64.1 & $ 45.66 \pm   1.23$ & $162.02 \pm   8.44$ & $354 \pm 42$ & $225 \pm 224$ & $1.20 \pm 0.21$ & $ 0.32 \pm  0.14$ & $ 1.17 \pm  0.15$ & $451 \pm 56$  \\
 79.9 & $ 51.09 \pm   1.94$ & $172.42 \pm  10.72$ & $339 \pm 52$ & $347 \pm 301$ & $1.28 \pm 0.46$ & $ 0.18 \pm  0.03$ & $ 1.65 \pm  0.25$ & $366 \pm 29$  \\
 98.7 & $ 56.53 \pm   3.12$ & $194.84 \pm  15.52$ & $291 \pm 74$ & $384 \pm 286$ & $1.11 \pm 0.64$ & $ 0.12 \pm  0.04$ & $ 2.53 \pm  0.63$ & $321 \pm 29$  \\
124.6 & $ 81.86 \pm   5.88$ & $266.15 \pm  23.66$ & $370 \pm 70$ & $-286 \pm 474$ & $1.12 \pm 0.67$ & $ 0.15 \pm  0.03$ & $ 5.67 \pm  2.66$ & $324 \pm 38$  \\
184.7 & $123.86 \pm   9.07$ & $593.94 \pm  44.55$ & $476 \pm 72$ & $626 \pm 327$ & $0.79 \pm 0.54$ & $ 0.24 \pm  0.05$ & $ 1.49 \pm  0.16$ & $456 \pm 40$ 
\enddata
\tablecomments{Same as table \ref{tab:m2lngals} for clusters binned by  \lvir.}
\end{deluxetable*}

\subsection{Corrections for Edges and Holes} \label{sec:edge}

\begin{figure}[t] \centering
    \includegraphics[scale=0.5]{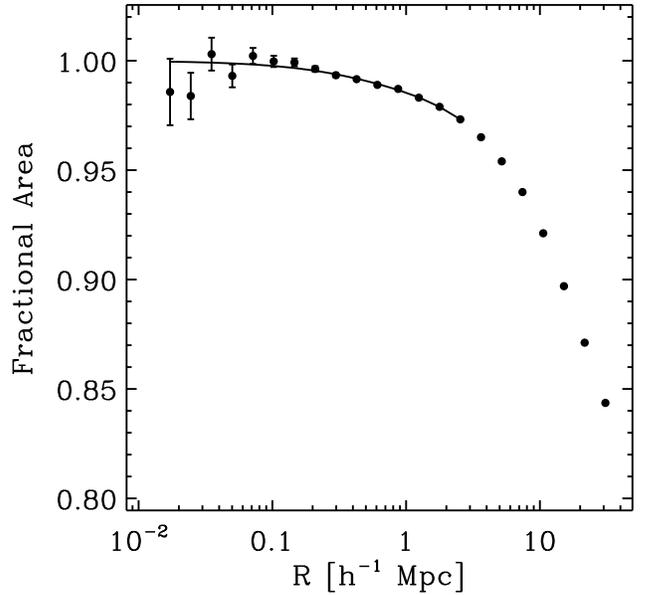}
    \caption{Mean fractional area searched relative to the total area in 
    each radial bin for the 8th cluster richness sample (\nvirsampdef).  Edges
    and holes dilute the true galaxy counts, biasing the measured density.
    This effect is negligible on small scales, but becomes important for large
    separations when a higher fraction of clusters are closer to the edge than
    the search radius.  Due to small area at small separations, the value is
    not well determined, but must approach unity smoothly.  We model this
    with a polynomial constrained to be unity on small scales, shown as the
    solid curve.  For larger scales no model is needed. \label{fig:usedarea}}

    \vspace{2em}
\end{figure}

The terms $D_p R_s$ and $R_p R_s$ introduced in \S \ref{sec:clmethods} correct
for the survey edges and holes by measuring the actual area searched.  An
example DR is shown in Figure \ref{fig:usedarea}, generated for one of the
richness bins described in \S \ref{sec:clusterbin}.  This is expressed as the
mean fractional area searched relative to the area in the bin. For small
separations, edges and holes make little difference so the fractional area is
close to unity.  On larger scales edges are important.   

\looseness+1 On small scales the number of pairs in each bin is relatively
small so the correction is not as well constrained.  However, we know that the
fractional area must approach unity smoothly, and this can be seen from visual
inspection.  To smooth the result, we fit a fifth order polynomial, constrained
to be unity on small scales, to the fractional area as a function of the
logarithmic separation.  Due to the weighting, this results in a curve that
approaches unity smoothly on small scales, yet matches intermediate separation
points exactly.  Points on larger scales are well-constrained and do not need
smoothing.

\section{Luminosity Density at z=0.25} \label{sec:ldens_universe}

We are interested in the background luminosity density at z=0.25, in the
$^{0.25}i$ band, for comparisons with our luminosity measurements.  We use an
evolved version of the SDSS spectroscopic sample (whose median redshift is
about $z\sim 0.1$). For this purpose, we use the DR4 version of the New York
University Value-Added Galaxy Catalog (NYU-VAGC; \citealt{blanton05a}). We
select a subset of the galaxies in the redshift range $0.01 < z < 0.25$, in the
apparent magnitude range $14.5 < m_r < 17.6$. The NYU-VAGC provides the angular
completeness map necessary to calculate for each galaxy the quantity
$V_{\mathrm{max}}$, the maximum volume over which each galaxy could be
observed.  We do so using the same method as used by \citet{blanton06a}, which
accounts for the evolution and $K$-corrections within this redshift range. We
use the {\tt kcorrect} {v4\_1\_4} code of \citet{blanton07a} to estimate the
$^{0.25}i$ band, Galactic extinction-corrected, $K$-corrected absolute
magnitude of each galaxy, based on the model fluxes provided by SDSS.

Finally, we evolution-correct these magnitudes in the following way, based on
the results of \citet{blanton06a}. For each galaxy we apply a simple
correction of the form:
\begin{equation}
	M_{^{0.25}i}(z=0.25) = M_{^{0.25}i}(z) + A (z-0.25)
\end{equation}
For galaxies on the blue sequence, based on their $^{0.25}(g-r)$ colors, we use
$A=0.65$.  For galaxies on the red sequence we use $A=0.35$.  The red-blue
split is defined by the line $^{0.25}(g-r) = 1.2 - 0.05 (M_{^{0.25}r} + 20)$.
These values are calculated by evaluating the simple star-formation history
models of \citet{blanton06a}. These models explain the evolution of the blue
and red sequences in the $^{0.1}g$ band well, and we use the corresponding
predictions for the $^{0.25}i$ band. In practice, these corrections are quite
small (at most 0.16 mag), and so the inevitable uncertainty in this correction
is likely to be unimportant.

By using the evolution-corrected magnitudes and weighting each galaxy by
$1/V_{\mathrm{max}}$, we estimate the luminosity function using the method of
\citet{schmidt68a}.  The resulting luminosity density for galaxies above our
luminosity threshold \lthresh\ is $(1.61 \pm 0.05) \times 10^8 h$
$L_\odot$Mpc$^{-3}$ in comoving coordinates, or \ldens\ in physical
coordinates.  Our luminosity limit is equivalent to \mshift\ $<$ \mthresh,
where we use $M_\odot = 4.67$ in the $^{0.25}i$ band. The uncertainty is
dominated by the absolute calibration of the SDSS. Fitting a single Schechter
function to the luminosity function, we find that $M_{\ast} - 5 \log_{10} h
\approx -20.9$ (with $\alpha \approx -1.21$), equivalent to $L_{\ast} = 1.7
\times 10^{10} h^{-2} L_\odot$.  Thus the lower luminosity threshold 
corresponds to 0.19$L_*$.

\section{Simulations} \label{sec:mocks}

In order to study the impact of the \maxbcg\ algorithm on our conclusions, and
in particular the differences between how our method operates on dark matter
halos and on \maxbcg\ clusters, we have repeated the luminosity measurements on
a mock catalog. These catalogs, which have been used in previous \maxbcg\
studies \citep{RozoCosmo07,RozoTool07,KoesterAlgorithm07,JohnstonLensing07},
populate a dark matter simulation with galaxies using the ADDGALS technique,
to be described in \citet{WechslerMock08}.  This method is designed to populate
large volume simulations with galaxies that have realistic luminosities,
colors, clustering properties, and galaxy clusters.

The catalog is based on the light-cone from the Hubble Volume simulation
\citep{Evrard02}, and extends from $0 < z < 0.34$.  Galaxies are assigned
directly to dark matter particles in the simulation, with a
luminosity-dependent bias scheme that is tuned to match local clustering data.
First galaxy luminosities are generated in the z=0.1 shifted $r$-band, drawn
from the luminosity function of \citet{BlantonLum03}.  The luminosity function
is assumed to evolve passively, with 1.3 magnitudes of evolution in $M_*$ per
unit redshift $( M_*(z) = M_*(z=0.1) - 1.3(z-0.1) )$.  Particles in the
simulation are then assigned these luminosities based upon the following
prescription.

We measure the local mass density around each dark matter particle, defined
here as the radius enclosing a mass scale of $\sim 1 \times 10^{13} M_\odot$.
For sets of points binned by local mass density, we measure the
auto-correlation function and the distance to the 5th nearest neighbor with the
same local density.  From galaxy surveys we know the correlation function as it
depends on galaxy luminosity, so by finding the set of particles with a
correlation function that matches that of galaxies in the real universe, we
make a connection between the local mass density in the simulation and galaxy
luminosity.  We use this to parameterize the probability distribution of these
dark matter densities as a function of luminosity, and constrain these
parameters so that the resulting luminosity-dependent two-point clustering
properties of the mock galaxies are in agreement with those measured in the
SDSS \citep {Zehavi05}.  

Once placed on a dark matter particle according to this prescription, each mock
galaxy is then assigned to a real SDSS galaxy that has approximately the same
luminosity and local galaxy density, measured here as the distance to the 5th
nearest neighbor.  The color for each mock galaxy is then given by the SED of
this matched galaxy transformed to the appropriate redshift.  The matching of
local galaxy density helps to ensure the relationship between color and density
is preserved.  

This procedure produces a catalog which matches several statistics of the
observed SDSS population, including the location, width and evolution of the
ridgeline in color-luminosity characteristic of galaxy clusters.  The
luminosity limit for galaxies in these mocks is slightly lower that of the
\maxbcg\ cluster finder $L > $\lstarlim, so the catalogs are well-designed for
testing the \maxbcg\ algorithm. However, this limit is higher than for our
luminosity measurements, so the results cannot compared at low luminosities.
Therefore, in this paper the simulations are used strictly to understand the
effects of cluster selection on our measurements (see \S \ref{sec:lumfunc}).

\section{Notation}

The notation may get confusing due to the use of multiple methods and apertures
in the course of cluster finding and lensing measurements.  The basic notation
for cluster variables, introduced in \S \ref{sec:cluster_sample}, is the same
as \citet{SheldonLensing07}: the measures of richness and luminosity we will
refer to as \nvir\ and \lvir.  These are the counts and i-band luminosity for
galaxies with $L_i > 0.4 L_*$, colors consistent with the cluster ridge-line,
and projected separation less than \rvirgal\ as calibrated in \citet{Hansen05}.
Note, \rvirgal\ was only used for \nvir\ and \lvir, no other quantities in this
paper use that aperture.  For more information about the richness measures see
\citep{KoesterAlgorithm07,KoesterCatalog07}.  This radius is different from the
radius \rvirmass\ calculated from the mass profile, which is typically half as
big \citep{JohnstonLensing07}; this difference is primarily due to differences
in convention between \citet{Hansen05} and \citet{JohnstonLensing07}: Hansen
used projected over-densities relative to the mean luminosity density and
Johnston used 3D over-densities relative to the critical mass density.  The
\lvir\ is only used for binning the clusters; because our results are
essentially the same for \lvir\ and \nvir\ binning we will avoid using the
\lvir\ notation except where necessary.  

We will refer to the total excess luminosity measured below, which includes the
light of all types of galaxies above a luminosity threshold, as \deltal$(r)$.
This luminosity, and excess mass \deltam$(r)$, are the new measurements
presented in this paper.  The total excess mass and light within \rvirmass\ are
denoted \deltamvir\ and \deltalvir.  Projected 2D radii are referred to as $R$
and 3D radii are referred to as $r$.

\section{Results} \label{sec:results}

In the following sections we show the results for clusters binned by \nvir.
Similar results were obtained binning by \lvir, and we summarize all results in
Tables \ref{tab:m2lngals} and \ref{tab:m2llum}, but for the sake of brevity we
include plots only for the \nvir\ binning.

\subsection{The Radial Dependence of the Joint Luminosity-color function}
\label{sec:lumcolorfunc}

Figure \ref{fig:lumcolor_vs_rad} shows the joint color-luminosity distribution
function for each radial bin in the 8th cluster richness sample
(\nvirsampdef).  Similar distributions were created for each of the cluster
richness bins used in this study.  We will present detailed analyses of these
type of data in \citet{Hansen07}, but we also present a sample here in order to
demonstrate a few aspects important for the M/L study.

The first point is that the population is quite different at small scales
relative to large scales.  On small scales, near the BCGs, the galaxy
population is dominated by red galaxies, while on large scales the color
distribution looks more like the cosmological average.  Similarly, on large
scales the luminosity function looks more like that of the average, although
they are poor fits to Schechter functions on large scales because we measure
the ratio of correlation functions (see the following sections for more
details).  These facts are relevant to the M/L for a few reasons.  We want to
make sure that the population we are seeing around clusters makes sense; that
we, for example, are complete in the color-luminosity range we have probed.
Again, detailed analysis will come in \citet{Hansen07}, but these color and
luminosity trends are exactly what is expected.  We will discuss the shape of
the luminosity function in \S \ref{sec:lumfunc}.

\subsection{The Radial Dependence of the Luminosity function}
\label{sec:lumfunc}

\begin{figure*}[p]

    \plotone{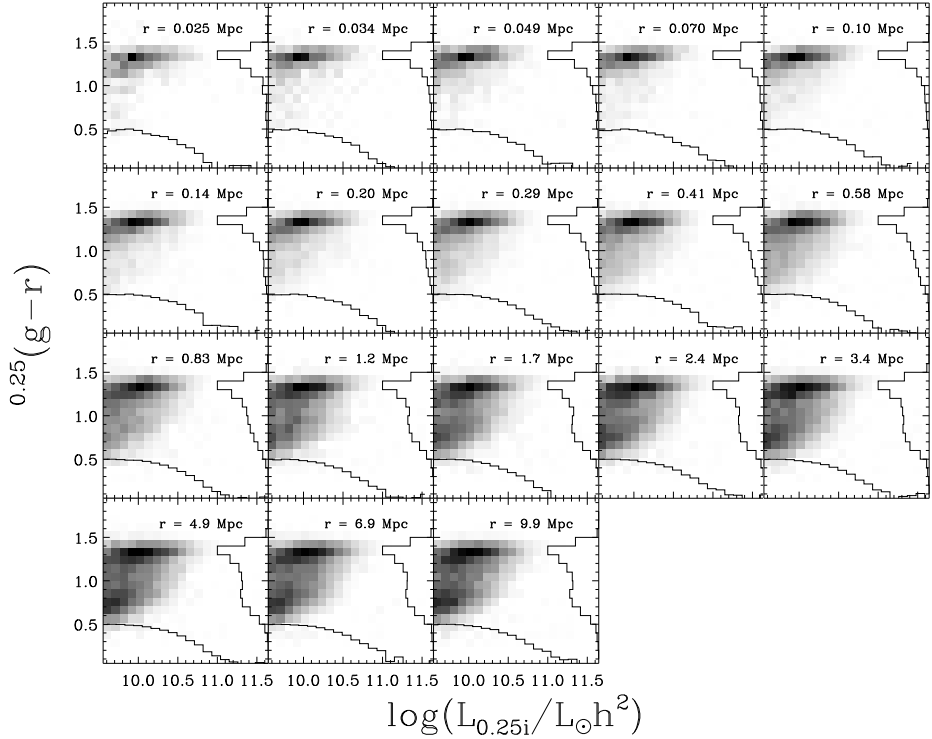}
    \caption{Joint galaxy \gmrshift\ and \ishift-band luminosity densities as a
    function of projected separation from BCGs for the 8th cluster richness bin
    (\nvirsampdef). The luminosities are expressed in the z=\zmean\ shifted
    bandpass.  Each frame corresponds to a different radial bin; the radius is
    indicated in the legend.  The one dimensional distributions for color and
    luminosity are also shown as the solid histograms along the left and bottom
    axes. The luminosity distribution is expressed as log of the number density
    as a function of log luminosity; the color distribution is linear density
    as a function of color.  At small separations red galaxies dominate while
    on large scales there is a bivariate color distribution similar to the
    cosmological average.  A smaller fraction of galaxies is highly luminous at
    small separations as compared to large.  \label{fig:lumcolor_vs_rad} }
     
\end{figure*}

Projecting the two-dimensional color-luminosity plots from Figure
\ref{fig:lumcolor_vs_rad} across the color axis produces the luminosity
function in each radial bin.  Recall that to produce these curves we have
statistically subtracted the background, so the luminosity function is the
luminosity function of galaxies minus that
mean density of the universe. The value of these functions in a
luminosity bin $i$ is
\begin{equation}
    \phi_{c,i}(R) = n(R)_i - \bar{n_i} = \phi_i w_{c,i}(R),
\end{equation}
where $n$ is the number density, $w_{c,i}(R)$ is the projected
cross-correlation function between clusters and galaxies in luminosity bin $i$
at projected radius $R$ and $\phi_i$ is the value of the luminosity function of
the universe in that luminosity bin $i$.  

\begin{figure*}[p] \centering

    \plotone{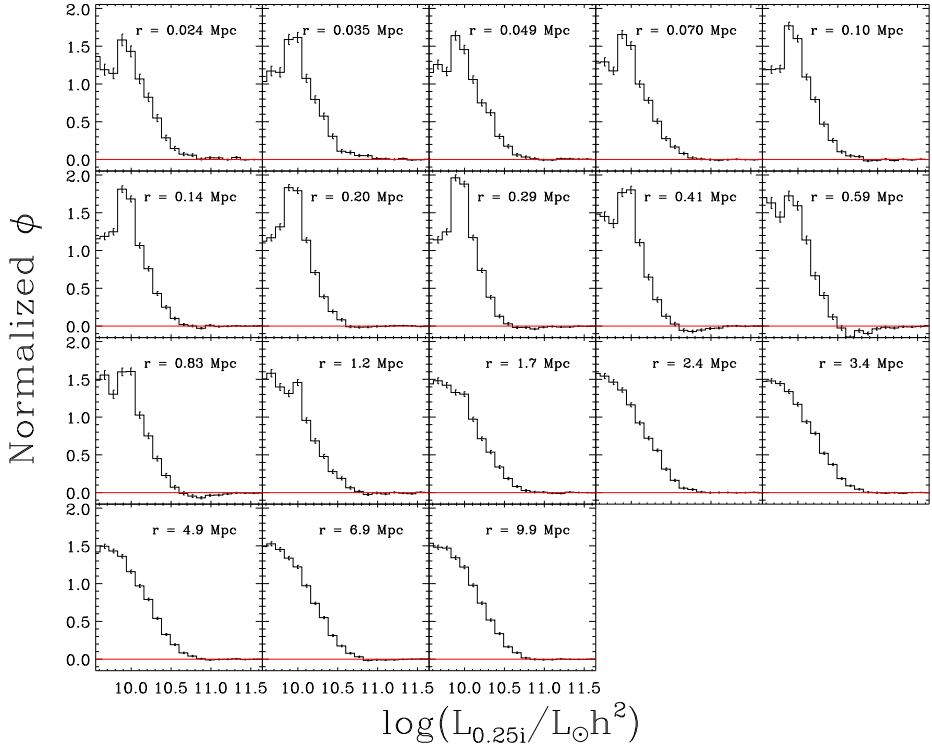}

    \caption{Excess luminosity functions for the first cluster richness bin
    \nvir$=3$.  This is the integral across the color axis of the joint
    densities such as those shown for another richness bin in Figure
    \ref{fig:lumcolor_vs_rad}.  As in that Figure, each panel is a radial bin
    with mean radius indicated in the legend.  Each bin in the excess
    luminosity function is the amplitude of the projected cluster-galaxy
    cross-correlation function at that radial separation times the mean
    density.  In some bins the galaxies are anti-correlated. This cluster
    richness bin was chosen as the extreme example of these anti-correlations;
    they are smaller or non-existent in higher richness bins.  As discussed in
    \S \ref{sec:lumfunc}, this feature is a result of the \maxbcg\ selection
    function at low richness.  }

\label{fig:lumfunc_vs_rad}
\end{figure*}

Because of the statistical subtraction, the value of the luminosity function in
luminosity bin $i$ is weighted by the cross-correlation function of clusters
with galaxies of that luminosity.  This is important because galaxies of
different luminosities correlate with clusters differently.  An extreme example
of this is demonstrated for the lowest \lvir\ bin in Figure
\ref{fig:lumfunc_vs_rad}, which shows $\phi_{c,i}$ in each radial bin.  This
Figure shows that, near the virial radius, the inferred luminosity function for
small groups is actually negative for galaxies with \lshift$\sim 10^{10.7}
L/L_\odot$.  This is not because there is a negative number density of these
galaxies around the groups, but because they are anti-correlated with the groups
at this separation.  This means that, near the virial radius, there are fewer
of these high luminosity galaxies relative to the background.  

\epsscale{0.80}

\epsscale{1.0}

This effect is strongest in our lowest \nvir\ bin, although there is a slight
feature in the luminosity function at virial radius for high luminosity galaxies
the first few cluster bins.  

In order to understand whether these effects are physical, due to selection
effects of the cluster finder, or are artifacts of our method, we ran the
\maxbcg\ algorithm and our cross-correlation code on the simulation-based mock
catalogs described in \S \ref{sec:mocks}.  We performed this test twice,
stacking on both cluster centers (BCGs) and halo centers.  For the mock stacked
on maxBCG clusters, a similar effect is seen, although it is suppressed
relative to the effect in the real data.  As in the data, the effect is
strongest in the lowest richness bin.  When the measurement is done around halo
centers,  there is no significant effect seen.  This indicates that it is
mostly an effect introduced by the selection criteria of the \maxbcg\
algorithm.

This lowest luminosity bin is peculiar in that it requires the close proximity
of only three very luminous red galaxies, one of which has extreme BCG-like
luminosity.  This an unusual situation; BCGs of this luminosity are usually
surrounded by many more galaxies.  In order to find only two such galaxies
within a few hundred kpc of a BCG type galaxy, the algorithm selects objects
embedded in slight under-densities.  This can occur naturally in the \maxbcg\
algorithm due to the percolation step, which does not allow clusters to be
embedded within larger clusters.  This may limit low richness systems to very
particular regions of space.

\subsection{Radial Luminosity and Number Density} \label{sec:radlum}

\begin{figure} \centering
    \plotone{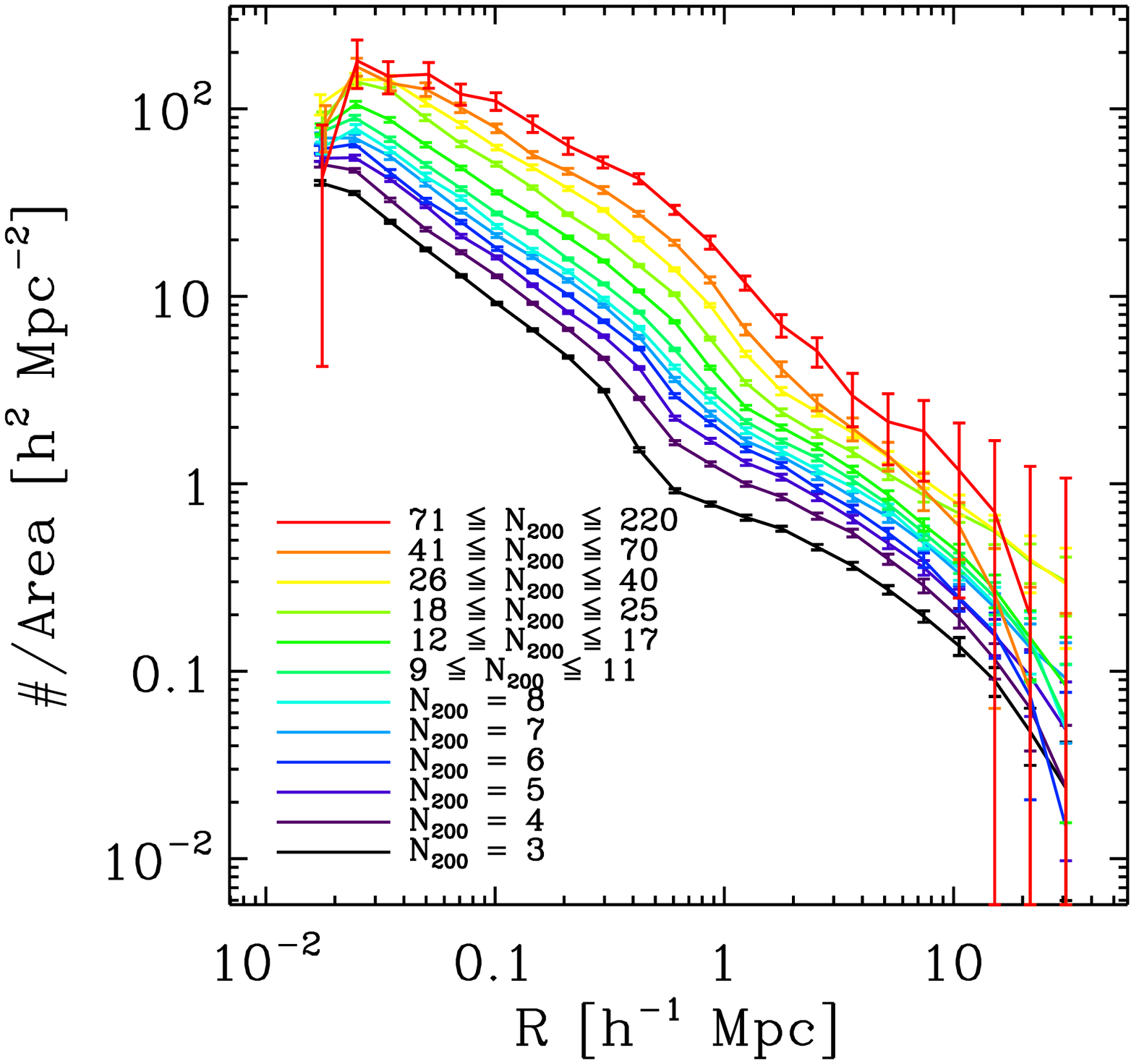}
    \caption{Excess radial number density for each of the cluster \nvir\ bins.
    These curves are the integral of the excess luminosity functions as shown
    in Figure \ref{fig:lumfunc_vs_rad}.}

    \label{fig:numdens}
    \vspace{2em}
\end{figure}

\begin{figure} \centering
    \plotone{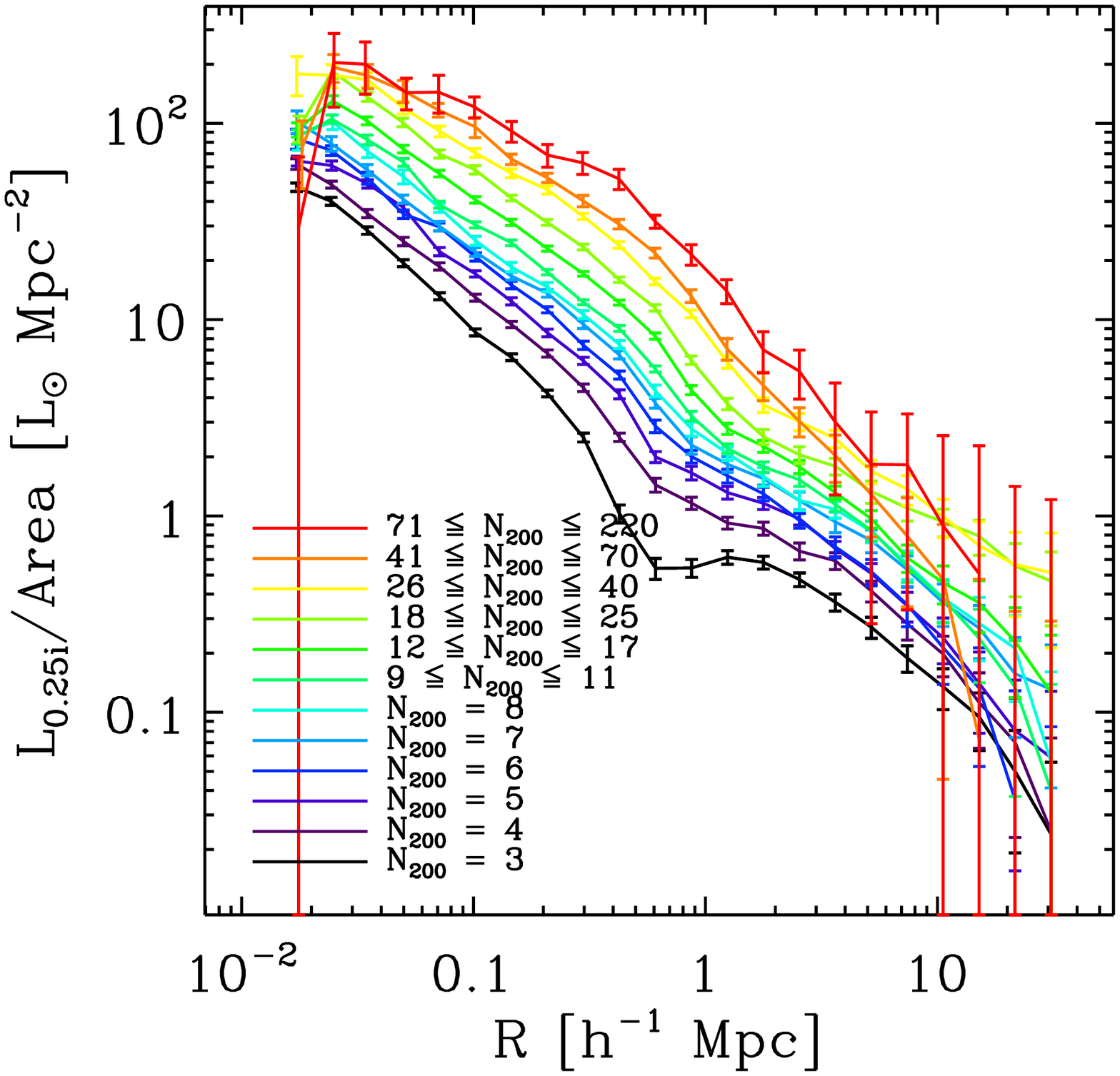}
    \caption{Excess radial \ishift-band luminosity density for each of the
    cluster \nvir\ bins. These curves are the integral of luminosity-weighted
    luminosity functions.}

    \label{fig:lumdens}
    \vspace{2em}
\end{figure}

Further integrating the luminosity functions from section \ref{sec:lumfunc}
across the luminosity axis results in the excess number density of galaxies.
We have also generated luminosity density profiles using the luminosity
weighted data cube rather than the cube of galaxy counts.   These are shown in
Figures \ref{fig:numdens} and \ref{fig:lumdens} respectively.  Each curve
represents the excess density for each of the cluster \lvir\ bins.  The errors
come from jackknife re-sampling of the data following the same techniques used
for the lensing analysis described in \citet{SheldonLensing07}.  We will say
more about errors in section \ref{sec:errors}.

Again, it is important to remember that these curves are 
background subtracted, and are thus the number and luminosity densities
above the mean.  These terms can be written in terms of correlation functions:
\begin{eqnarray} \label{eq:density_projected}
n(R) - \bar{n} & = & \bar{n}~w_{c,g}(R) \\
\ell(R) - \bar{\ell} & = & \bar{\ell}~w_{c,\ell}(R) 
\end{eqnarray}
where $\bar{n}$ and $\bar{\ell}$ are the mean number and luminosity density of
galaxies over the explored luminosity range, and $w_{c,g}(R)$ and
$w_{c,\ell}(R)$ are the projected cluster-galaxy and cluster-luminosity
cross-correlation functions, averaged over all galaxies in the luminosity range
at projected radius $R$, weighted by the luminosity function.

It is tempting to interpret the ratio of the excess luminosity to excess number
densities as the mean luminosity as a function of radius. Given that there can
be negative densities with respect to the mean at some luminosities when there
are anti-correlations, as shown in Figure \ref{fig:lumfunc_vs_rad},  this is
not always the correct interpretation.

The curves in figures \ref{fig:numdens} and \ref{fig:lumdens} show a number of
features expected for two-point correlation functions.  In particular, there
should be a transition radius between correlations within the halo and between
different halos.  The scale of this break should correspond to the size of the
larges clusters in each bin.  We present no detailed analysis here, but the
radius of the break we see does increase with cluster richness as expected.  On
small scales the profile is consistent with a universal profile and on large
scales transitions to that expected for halo-halo correlations.  We found in
\citet{JohnstonLensing07} that the mass profiles of the clusters were good fits
models of this form.  In this paper we do not use any explicit modeling,
preferring to focus on model independent measurements, but these rich data
should provide constraints for models of galaxy formation and evolution in a
cosmological setting.

\subsection{Integrated Luminosity Profiles} \label{sec:lum_profiles}

We will use integrated luminosity profiles to compute the mean M/L within a
given three-dimensional radius $r$.  We invert the projected two-dimensional
profile shown in Figure \ref{fig:lumdens} using a standard Abel type inversion
\citep*[e.g][]{Plummer1911}: 
\begin{equation} \label{eq:lum_density}
    \Delta \ell(r) = \ell(r) - \bar{\ell} = \frac{1}{\pi} \int_{r}^{\infty} \mathrm{d}R \frac{-\ell^{\prime}(R)}{\sqrt{r^2 - R^2}},
\end{equation}
where we have re-used the notation for the excess luminosity density from
equation \ref{eq:density_projected} but replaced projected radius $R$ with
three-dimensional radius $r$.  We have been explicit here in indicating we
measure the density minus the mean $\ell(r) - \bar{\ell}$.  Thus the $\Delta
\ell(r)$ is proportional to the cluster-light cross-correlation function:
\begin{equation}
    \Delta \ell(r) = \bar{\ell}~\xi_{c\ell}(r)
\end{equation}

The assumption behind the inversion in equation \ref{eq:lum_density} is that
$\ell(R)$ is the line of sight projection of a spherically symmetric
three-dimensional function $\ell(r)$.  This follows from the isotropy of the
universe as long as the cluster selection function does not select structures
preferentially aligned relative to us.  Results from simulations suggest
such preferential alignment is not important for our cluster sample where
the sample can be tested \nvir$>10$.  A publication on these simulations
is forthcoming \citep{Hao08}.

Because the maximum separation we measure is $30 h^{-1}$ Mpc rather than
infinity, we cannot accurately calculate $\Delta \ell(r)$ over the entire
range.  We lose the last point entirely, and the second to last point, at
\maxinvertrad, must be corrected slightly for the endpoint.  We perform a
power-law extrapolation of the profile and find this to be a 5\% upward
correction.

\begin{figure}[t] 

    \centering \epsscale{0.95}
    \plotone{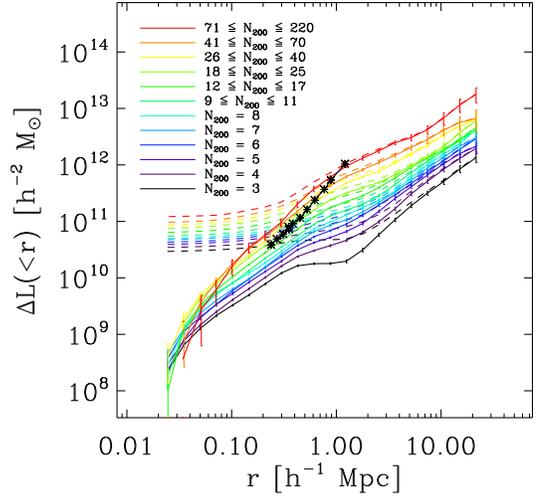}
    \caption{Excess 3D \ishift-band luminosity profiles for each of the cluster
    \nvir\ bins.  The 2D profiles from Figure \ref{fig:lumdens} were
    de-projected and integrated over radius.  The dashed curves without error
    bars include the BCG luminosity, the solid lines with errors do not.  The
    mean \rvirmass, determined from the mass profiles in \S
    \ref{sec:mass_profiles}, is marked with an asterisk in each curve.}
    \label{fig:ngals200_lumboth} \vspace{2em}

\end{figure}

We then integrate the excess luminosity density to obtain the total luminosity
within radius $r$.  Because the profile only extends inwards to \minrad, we are
missing some light interior to this radius.  This light is dominated by the BCG
however, so we can add that component back in as the average BCG luminosity for
a given cluster sample:
\begin{equation}
    \Delta L(<r) = \langle L_{BCG} \rangle + \int_{rmin}^{r} \mathrm{d}r 4 \pi r^2 \Delta \ell(r),
\end{equation}
where we have again been explicit in defining our measured quantity as $\Delta
L(<r)$, the integrated excess luminosity above the mean density. 

Figure \ref{fig:ngals200_lumboth} shows this quantity $\Delta L(<r)$ for the
richness and luminosity cluster binning.  There are two curves for each
richness bin: one including the mean BCG luminosity $\langle L_{BCG}\rangle$
and one without.  The light is dominated by the BCG on small scales.  The
luminosity within \rvirmass\ is marked with an asterisk.  The \rvirmass\ is
calculated from the mass profiles (see \S \ref{sec:mass_profiles}).

\newpage
\subsection{Integrated Mass Profiles} \label{sec:mass_profiles}

The cluster mass profiles were measured from the lensing measurements presented
in \citet{SheldonLensing07} and \citet{JohnstonLensing07}.  These measurements
were performed for the same samples presented above. The basic lensing
measurement is \deltasig, which is a projected quantity:
\begin{equation} \label{eq:deltasig}
\Delta \Sigma \equiv \bar{\Sigma}(< R) -\bar{\Sigma}(R) ~,
\end{equation}
where $\bar{\Sigma}$ is the projected surface mass density at radius $R$ and
$\bar{\Sigma}(<R)$ is the mean projected density within radius $R$.  The
subtraction in this equation is a manifestation of the mass sheet degeneracy.

In \citet{JohnstonLensing07} we inverted the projected profiles to the
three-dimensional excess mass density using the techniques presented in
\citet{JohnstonInvert07}. This inversion is a procedure analogous to the
inversion of the luminosity density presented in \S \ref{sec:lum_profiles}:
\begin{eqnarray} \label{eq:invert}
    -\frac{\mathrm{d}\Sigma}{\mathrm{d}R} & = & \frac{\mathrm{d}\Delta\Sigma}{\mathrm{d}R} + 2\frac{\Delta\Sigma}{R} \nonumber \\
    \Delta \rho(r) & \equiv & \rho(r) - \bar{\rho} = \frac{1}{\pi} \int_r^{\infty} \mathrm{d}R \frac{-d\Sigma/\mathrm{d}R}{\sqrt{R^2-r^2}}.
\end{eqnarray}
Again, the assumption is that the profiles are projections of spherically
symmetric three-dimensional functions.  We can recover the total excess
mass within radius $r$, including that within our innermost radius, because
\deltasig\ is itself a non-local measurement:
\begin{eqnarray} \label{eq:masstot}
\Delta M(<r) & = & \pi R^2 \Delta \Sigma(R)~ + \nonumber 
\end{eqnarray}
\begin{equation}
    2\pi\int_R^\infty \mathrm{d}r~r 
    \Delta \rho(r)\left[ \frac{R^2}{\sqrt{r^2-R^2}} - 2\left( r-\sqrt{r^2-R^2} \right) \right]
\end{equation}

\begin{figure*}[p] \centering
    \plotone{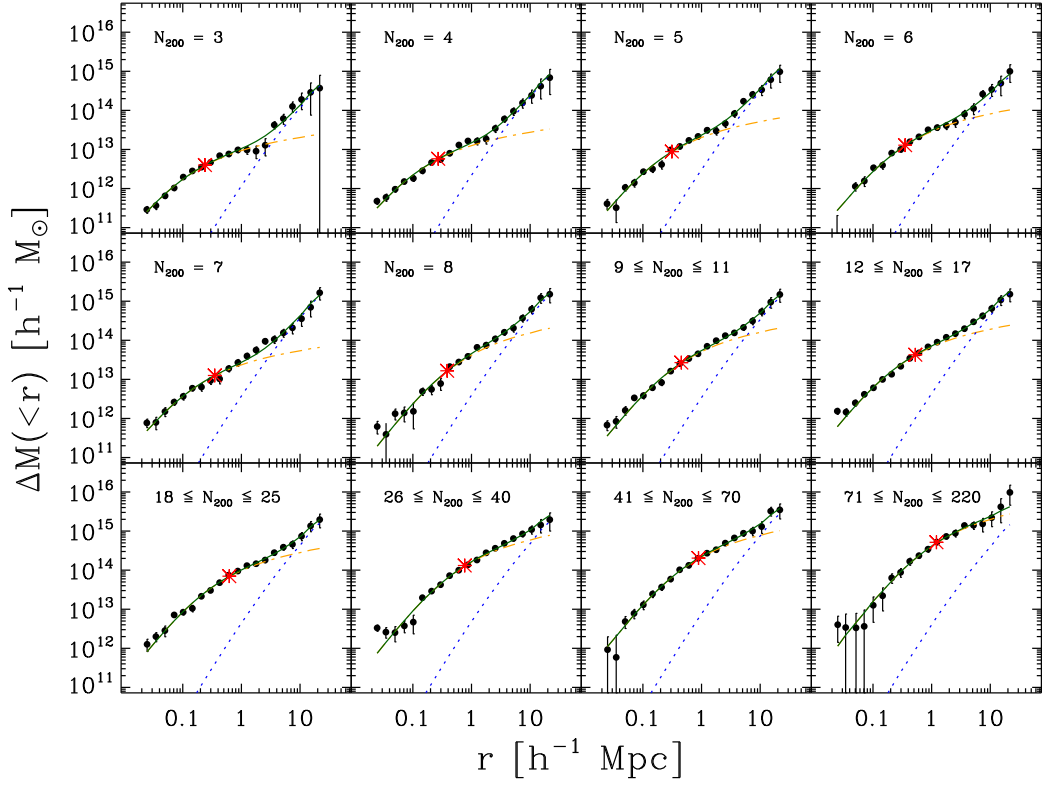}

    \caption{Cumulative 3D excess mass profiles for each cluster \nvir\ bin
    derived from weak gravitational lensing.  The solid curves show the best
    fit NFW+linear bias model. The dotted curve is the corresponding linear
    model and the dot-dashed curve is the NFW only.  The asterisk marks
    \rvirmass.  Note these models do not account for possible offsets between
    the BCG and the halo mass peak, which would alter the profile on small
    scales.}

    \label{fig:ngals200_mass} 
    
    \end{figure*}

As with the luminosity inversions, the last point must be thrown out and a 5\%
correction is applied to the second to last point.  Figure
\ref{fig:ngals200_mass} shows these excess mass profiles for the \nvir\
binnings.  The error bars are from jackknife re-sampling.  See
\citet{SheldonLensing07} for details about the error estimates, and \S
\ref{sec:errors} for more details about errors in this work. These data are the
same as presented in \citet{JohnstonLensing07}.  We note that there could be a
level of systematic error in these measurements, primarily from the photometric
redshift errors on the background source galaxies used for the shear
measurement.  It is difficult to know this error, but simulations suggest the
calibration is good to 10\%.

In order to get a size scale for each cluster sample, we fit a simple model to
the data.  The model is that of an NFW profile on small scales \citep{nfw97}
and linear correlations on large scales.  This model was presented in detail in
\citet{JohnstonLensing07}.  For this paper, we only use this fit in order to
estimate a size \rvirmass, from which we can also calculate the mass
\deltamvir.  The \rvirmass\ is the radius where the mean mass density falls to
200 times the critical density, and \mvir\ is the mass contained within that
radius.  This radius \rvirmass\ will be a reference point for the M/L
measurements.  The fits for each richness bin are shown in Figure
\ref{fig:ngals200_mass} and \rvirmass\ is marked for reference.  Note, these
values are somewhat different than those in \citet{JohnstonLensing07} where
power-law interpolation was used to extract \rvirmass.

In \citet{JohnstonLensing07} it is shown that if a fair fraction of the BCGs
are not centered on the peak of the halo mass distribution, the shape can be
strongly effected.  In that work an offset distribution was determined
from simulations, and used to recover the underlying halo mass distribution.
In this work we do not try to recover halo masses, but work directly with the
observations around the BCGs chosen as centers.  Thus it is important to keep
in mind that these are the mass profiles (and light profiles) around BCGs, not
necessarily peaks of the halo mass distribution.  This is probably not a large
effect for the M/L at the virial radius, and is negligible on the largest
scales. On the other hand, for relatively small radii the shape of the M/L
profile around BCGs may be different than that around halo centers.

\subsection{Mass-to-light Ratio Profiles} \label{sec:m2l_profiles}

\begin{figure*}[p] \centering
    \plotone{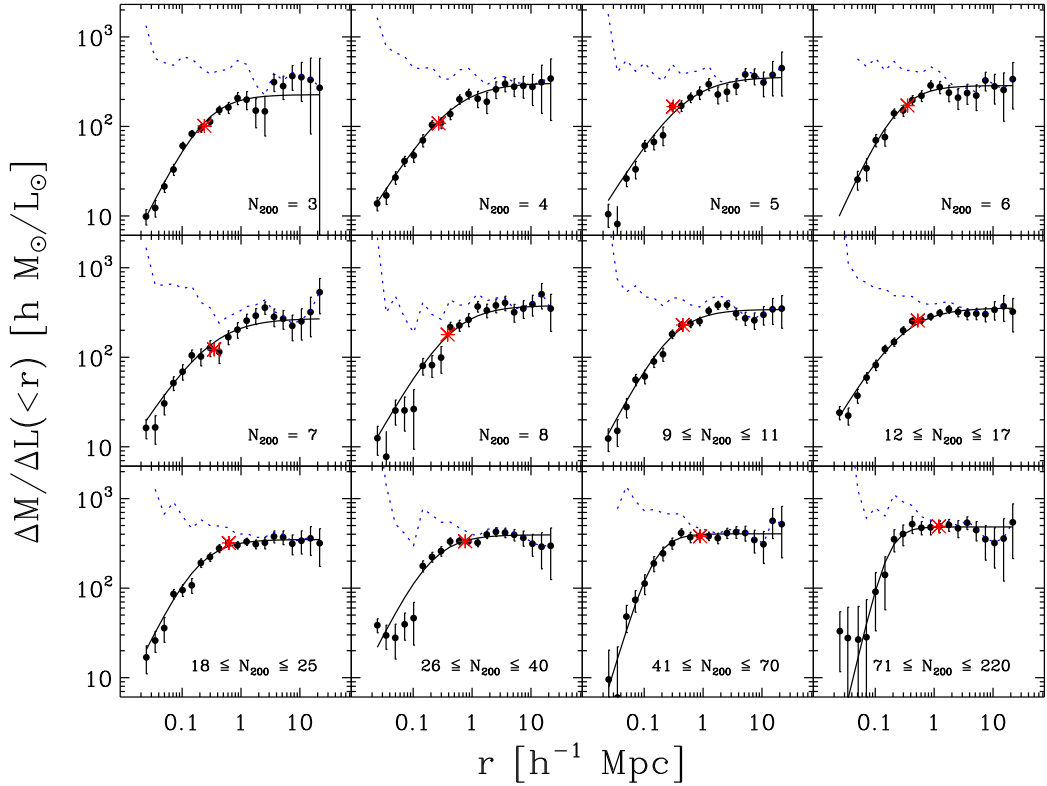} 
    
    \caption{Excess mass to excess light ratio profiles for each of the \nvir\
    bins.  Light is measured in the \ishift\ bandpass.  These curves are the
    ratio of the curves shown in Figures \ref{fig:ngals200_lumboth} and
    \ref{fig:ngals200_mass}.  The points with error bars include the mean BCG
    luminosity, while the dotted curves exclude the luminosity of the BCG.  The
    asterisk marks \rvirmass. The curve through the data is a simple
    descriptive model as discussed in \S \ref{sec:m2l_profiles}.  }
    \label{fig:ngals200_m2lfits} 
    
\end{figure*}

In order to generate \deltamtol\ profiles we simply divide the
integrated excess mass \deltam\ by the integrated excess light \deltal\
profiles.  These profiles are shown for each cluster richness bin in Figure
\ref{fig:ngals200_m2lfits}.

\epsscale{0.9} 

\epsscale{1.0}
The \deltamtol\ is shown with and without the mean BCG luminosity.  When
the mean BCG luminosity is included the profile rises steeply and then 
flattens out at large radius.  The \rvirmass\ is marked for reference. The
mean \deltamtol\ within \rvirmass\ is a strong function of cluster richness.
However, the asymptotic \deltamtol\ is nearly independent of cluster
richness.

Without the mean BCG luminosity included, the \deltamtol\ is relatively flat on
intermediate to large scales, indicating that the relative amount of mass and
light on those scales is not a strong function of radius.  There is a turn up
at small scales, however.  This is partly due to the fact that we do not
measure light on scales less than \minrad.  However, there is also light not
counted in our luminosity measurements.  It is known that there is a
significant amount of intra-cluster light (ICL), light not associated with
galaxies, in many clusters \citep*[e.g][]{GonzalesICL05}.  There is also
missing light from galaxies below the luminosity threshold \lthresh\ and the
light not counted from the outskirts of detected galaxies (probably dominated
by the outskirts of the BCG on these scales).  Only by estimating this missing
light will we know the true M/L profile on small scales, and the absolute M/L
for all excess light in these clusters.  

In order to more quantitively describe the shape of the \deltamtol\ curves, we
use a simple fitting function that captures the main features of our profiles.
It is a function which would describe the ratio of two equal-index power laws
in mass and light plus a delta function for the mean BCG luminosity.
\begin{eqnarray} \label{eq:m2lmodel}
\Delta M/\Delta L (<r) & = &  
\frac{M_0 (r/1 h^{-1}\textrm{Mpc})^\alpha}{\langle L_{BCG} \rangle + L_0 (r/1 h^{-1}\textrm{Mpc})^\alpha} \nonumber \\
    & = & 
    \left( \frac{ (r/\textrm{\rhalf})^\alpha}{1 + (r/\textrm{\rhalf})^\alpha}\right) \left(\frac{\Delta M}{\Delta L}\right)_{asym} , 
\end{eqnarray}
where \rhalf\ $ = (\langle L_{BCG} \rangle / L_0)^{1/\alpha}$ is the radius at
which the \deltamtol\ reaches half its asymptotic value at infinity \mtolasym.
A larger \rhalf\ at fixed $\alpha$ implies a larger fraction of the total light
is in the BCG, which results in a slower transition to the asymptotic
\deltamtol.

Although we know the mass and light profiles are not pure power laws, they tend
to deviate from a power law in similar ways, which partially cancels this error.
However, this means that $\alpha$ should not be interpreted as the slope of the
mass or light profile.

The best fits for equation \ref{eq:m2lmodel} are over-plotted in Figure
\ref{fig:ngals200_m2lfits} for the \nvir\ binning.  We use the full covariance
matrix of the \deltamtol, generated from the covariance matrices of \deltam\
and \deltal, for the fits.  Using the covariance matrix accounts for the strong
correlations in the errors caused by the radial integration. The best-fit
\rhalf, $\alpha$, and \mtolasym\ are listed in Tables \ref{tab:m2lngals} and
\ref{tab:m2llum}.  There is a weak trend of decreasing \rhalf\ with \nvir,
while $\alpha$ is relatively constant except for the highest richness bins.
Recall that \rhalf\ $ = (\langle L_{BCG} \rangle / L_0)^{1/\alpha}$, and thus a
smaller \rhalf\ at fixed $\alpha$ is indicative of a less dominant BCG relative
to the overall cluster luminosity, which we had already seen to be true in
Figure \ref{fig:ngals200_lumboth}.  However, at high masses the value of
$\alpha$ increases also, weakening the change in \rhalf\ somewhat.  We will
discuss the \mtolasym\ values in section \ref{sec:m2lasym}.

\subsection{Mass-to-light Ratio Within \rvirmass}

\begin{figure}[t] \centering
    
    \plotone{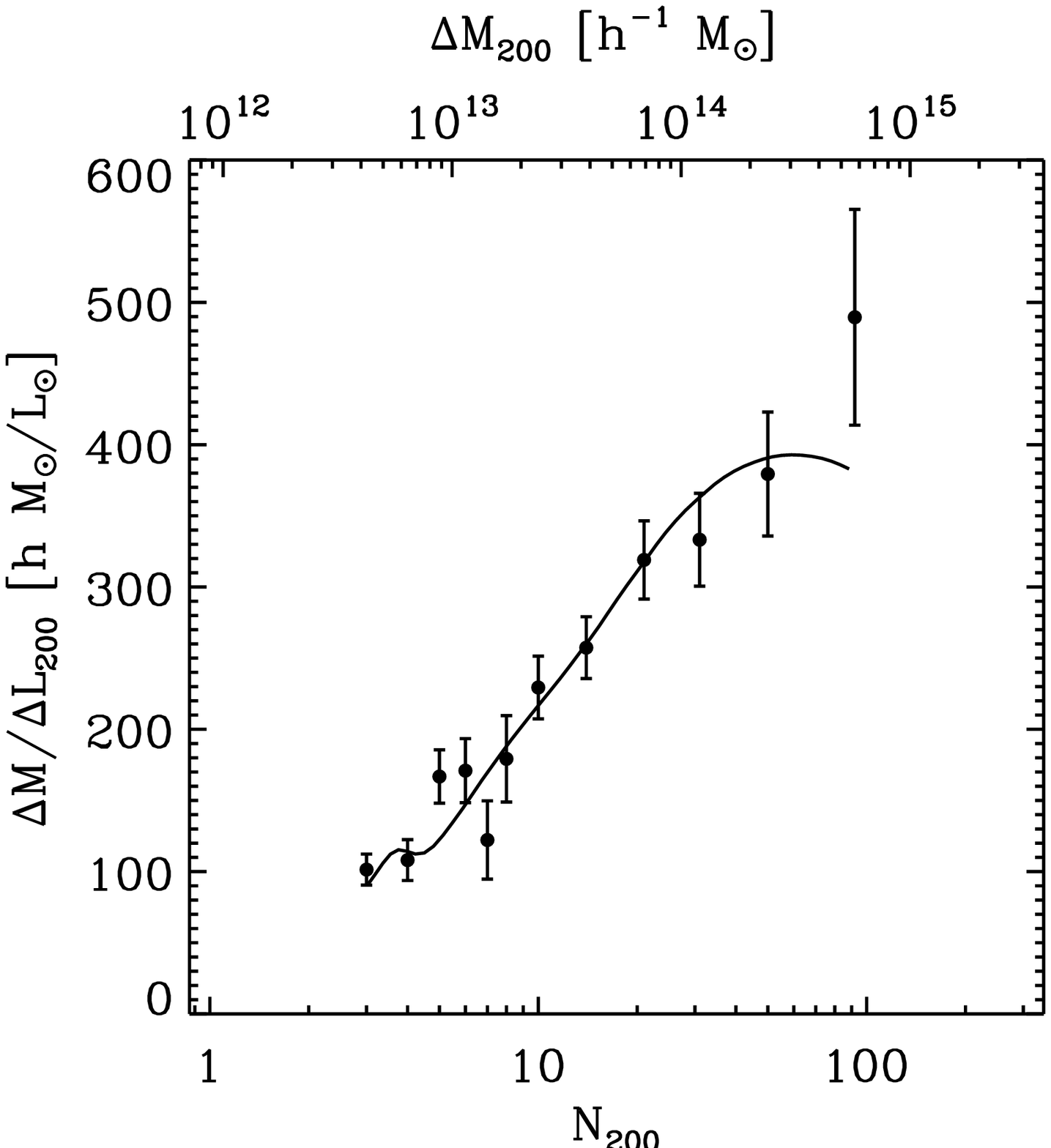} \caption{The
    mean \deltamtol\ within \rvirmass\ as a function of \nvir.  Light is
    measured in the \ishift\ bandpass.  The mean \deltamvir\ within \rvirmass\
    centered on BCGs is shown on the top axis, although this is only rough
    since the transformation between \nvir\ and mass is non-linear.  No
    correction for offsets between the BCGs and the halo mass peaks is
    included.  The \mtolvir\ vs \deltamvir\ is a good fit to a power law with
    index $0.33 \pm 0.02$; the \mtolvir\ vs \nvir\ is not a good power law.
    The over-plotted curve is a prediction based on the simulations of
    \citet{TinkerM2l05}. This curve is not correctly matched to this data, but
    is rather intended to show the rough expected trend with mass. }
    \label{fig:ngals200_m2l200} 

    \vspace{2em}
\end{figure}

In Figure \ref{fig:ngals200_m2l200} we show the mean \deltamtol\ within
\rvirmass, \mtolvir, for each of the cluster richness and luminosity bins,
plotted as a function of the mean \nvir.  The \deltamtol\ increases strongly
with \nvir\ over two orders of magnitude in mass.  The \deltamvir\ is also
shown on the top axis, but note this is rough as the transformation is
non-linear.  The \mtolvir\ versus \deltamvir\ is a good fit to a power law with
index $0.33 \pm 0.02$. The \mtolvir\ vs. \nvir, however, is not a good fit to a
power law due to the non-linear relationship between mass and galaxy counts.

Over-plotted in Figure \ref{fig:ngals200_m2l200} is a prediction based on the
models in \citet{TinkerM2l05} for $z=0$ rest-frame $i$-band light rather than
\ishift. The predicted \nvir\ have been scaled by a factor of 1.5 to pass over
the points. We do not expect this prediction to match our data, which is in a
different band and for which there are BCG centering effects. The point here is
to show a rough expected trend with mass.  There is qualitative agreement with
the mass scaling of this prediction and our data.  Note, it is tempting to
think our data do not asymptote at high mass as expected from the model, but
there is actually agreement at the one sigma level.

\subsection{Asymptotic Mass-to-light Ratio}\label{sec:m2lasym}

\begin{figure}[t] \centering
    \plotone{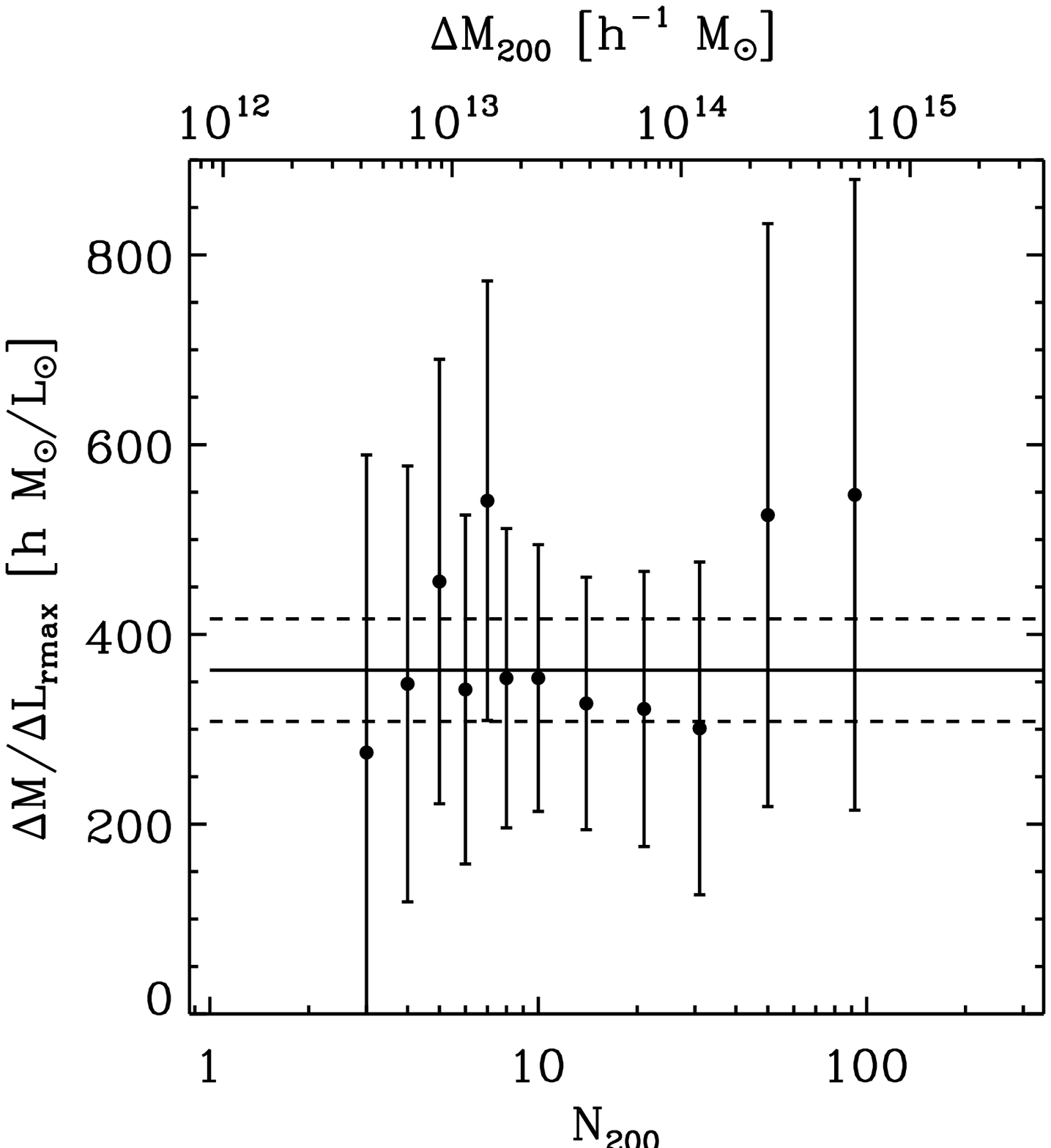}
    \caption{Asymptotic excess mass-to-light ratio as a function of \nvir.
        Light is measured in the \ishift\ bandpass.
    This is simply the last point on the integrated \deltamtol\ curve at $r = $
    \maxinvertrad.  The mean \deltam\ within \rvirmass\ centered on BCGs is
    shown on the top axis, although this is only rough since the transformation
    between \nvir\ and mass is non-linear. The mean \deltamtol$ = 362 \pm 54
    h$, averaged over all samples, is plotted as the horizontal line}
    \label{fig:ngals200_m2lasym} 

    \vspace{2em}
\end{figure}

The \deltamtol\ profiles shown in Figure \ref{fig:ngals200_m2lfits} rise
quickly and flatten at large separations.   We measure this asymptotic value in
two separate ways. First we use the last point in the \deltamtol\ curve at $r <
$\maxinvertrad, which we will refer to as \mtolmax.  The \mtolmax\ for each
richness bin is shown in Figure \ref{fig:ngals200_m2lasym}.  Note, the value at
\maxinvertrad\ should be insensitive to any offsets between the BCGs
and the true halo mass peak.  We detect no trend in the \mtolmax\ as a
function of \nvir.  For the average value we get \meanmtolrmaxn\ for the
richness binning and \meanmtolrmaxl\ for the luminosity binning.  Note these
are not independent measures since the same clusters are used for both
binnings.

For the second method we examine the \mtolasym\ values measured from the fits
in \S \ref{sec:m2l_profiles}, which are listed in  Tables \ref{tab:m2lngals}
and \ref{tab:m2llum}.  As with the \mtolmax\ values, they are roughly constant
with \nvir.  The errors on \mtolasym\ are much smaller than those on
\mtolmax\ because the model uses all data points rather than just the
last data point to infer the asymptotic mass-to-light ratio. We will discuss
the interpretation of these errors more in the discussion.  

The average model asymptotic mass-to-light ratios are \meanmtolasymn\ for the
richness binning and \meanmtolasyml\ for the luminosity binning.  Note the two
binnings are not independent measures as the same clusters are used for both.
Both are consistent with the values we get just taking the last point \mtolmax.

\section{Errors} \label{sec:errors}

As stated earlier, all errors come from jackknife re-sampling of the data.  The
method was discussed in \citet{Sheldon04,SheldonLensing07} in context of the
lensing measurements.  The same technique was used for the light measurements.
The main difference between jackknifing in the lensing and luminosity
measurements is that for the luminosity correlations we must jackknife all
pieces of the estimator in Equation \ref{eq:estimator}.  

The systematics in the M/L measurements are dominated by calibration
uncertainties in the lensing measurements, in particular the photometric
redshift estimates for the background source galaxies.  These errors were
discussed \S \ref{sec:lensmethods} and more detail can be found in
\citet{JohnstonLensing07}.  Although it is difficult to know the absolute scale
of the expected uncertainties, based upon the results of both simulations and
real world tests shown in \citet{Lima08} we estimate the overall level of
systematics to be of order 10\% for all contributing factors.

We want to stress again that the errors on the asymptotic mass-to-light ratio
differ greatly for the two methods for a very simple reason:  the \mtolmax\ is
derived from the last point on the M/L curve and the \mtolasym\ is derived from
fitting a simple descriptive model to {\it all points}.  Because this model is
not a physical model the points on small scales should not be expected to be
constraining of the large scale M/L.  So the errors on \mtolasym\ must be
thought of as lower limits at this stage.  The errors on \mtolmax\ can be
thought as upper limits since there is certainly some information in points at
smaller radius.

\section{Discussion}

The integrated \deltamtol\ around \maxbcg\ clusters has a generic form.  At
\minrad, where the light of the cluster is dominated by the BCG, the
\deltamtol\ is  $\sim 10 h^{-2} M_\odot/L_\odot$.  There is a sharp rise and
then the profile flattens out at large scales (\maxinvertrad).  We fit a simple
model to extract \rhalf, the radius at which the \deltamtol\ reaches half its
value at infinity (see \S \ref{sec:m2l_profiles} for details).

For \mvir$< 10^{14} M_\odot$, \rhalf\ is determined primarily by the relative
amount of light in the BCG compared to the rest of the cluster.  For higher
masses the \deltamtol\ asymptotes to \mtolasym\ more quickly relative to the
lower masses, resulting in a smaller \rhalf.  

The decreasing \rhalf\ with richness is partly due to the less dominant BCGs
for higher richness clusters; the effect of the BCG on the integrated light is
only important on small scales for very rich systems.  But it is also partly
because the measured \deltamtol\ within the large clusters is closer to the
universal value.  Figure \ref{fig:ngals200_m2lfits}  indicates that the
\deltamtol\ not including the BCG light is more flat for the high richness
clusters that have especially sharp transitions. The \deltamtol\ would be
essentially equal to \mtolasym\ if it were not for the presence of the BCG.
This is not true for the lower richness systems.

This difference between high and low richness systems leads us back to the
discussion of uncounted light.  Uncounted light is any light not counted in our
measurements.  This uncounted light is partly intra-cluster light (ICL), light
not associated with galaxies. There is also light from galaxies less luminous
than the threshold \lthresh. The total luminosity of galaxies below this
threshold is probably not dominant, and the radial profile is probably similar
to that of galaxies above the threshold, so including it would not change the
profile dramatically.  But the ICL has a steeper profile.  It appears to follow
a $\textrm{exp}(-r^{1/4})$ law, with scale length of order 100 $h^{-1}$ kpc,
and contains many times more light than the BCG \citep{GonzalesICL05}.  The
total light in this component scales slowly with richness, and is more dominant
on small scales in smaller systems. This could explain what appears to be a
slower rise in \deltamtol\ relative to larger systems.  We will explore the ICL
for \maxbcg\ clusters in a future paper.

The \mtolvir, the excess mass-to-light within \rvirmass, scales with richness
and \mvir.  For lower richness systems, the \mtolvir\ is considerably smaller
than \mtolasym, while for larger systems it is of order \mtolasym.  This trend
is probably a reflection of both a true difference in mean \deltamtol\ and the
effect of the ICL, which may be more dominant for lower mass systems.  The
\mtolvir\ versus \nvir\ is not a good fit to a power law, but \mtolvir\ versus
\deltamvir\ is well fit by a power law with index $0.33 \pm 0.02$ (there is a
non-linear relationship between mass and galaxy counts).  However, no attempt
was made to model possible offsets between BCGs locations and the halo mass
peaks. The \deltamtol\ will not be strongly affected because the luminosity
roughly traces mass, the \deltamvir\ as a function of \nvir\ does change
significantly \citep{JohnstonLensing07}.  A fully model-dependent analysis for
both the mass and light profiles following \citet{JohnstonLensing07} may imply
different results for \mtolvir\ measured around dark matter halos than those
around \maxbcg\ clusters.

It is difficult to compare with the literature due to the many conventions and
methods in use with regards to cluster selection, lower luminosity thresholds,
galaxy aperture definitions, mass apertures and estimators, projected vs.
de-projected masses, luminosities with or without background subtraction, and
the various bandpasses used for the exposures.  With these caveats, we will say
that there is broad agreement in the literature that $M/L \varpropto
M^{0.2-0.3}$ \citep[e.g,][]{GirardiM2l00,LinMohrStanford04,PopessoRassSdss07}.
Below we compare the $\Omega_m$ calculated from the inferred global
mass-to-light ratio, which may be less dependent on these factors.

We used two methods to extract the asymptotic \deltamtol: the value with in
\maxinvertrad, \mtolmax, and the best-fit value from our fitting function,
\mtolasym.    Note, on these large scales, any offsets between the BCGs
positions and the halo mass peak is irrelevant. We see no trend of either
measure of the asymptotic \deltamtol\ with \nvir. 

As we discussed in the introduction, for any given cluster sample, the
asymptotic \deltamtol\ is proportional to the mass-to-light ratio of the
universe.  Repeating equation \ref{eq:m2lbias_test} here for clarity:
\begin{eqnarray} 
    \textrm{\mtolasym} & = & \langle M/L \rangle~ \textrm{\bmtolinv} \nonumber \\
    \textrm{\bmtolinv} & = & \frac{b_{cm}}{b_{c\ell}} \frac{1}{b^2_{\ell m}}.\nonumber
\end{eqnarray}
where \bmtol\ depends primarily on the bias of the galaxy tracers relative to
the underlying mass distribution, since the bias of the clusters likely cancels
out in $b_{cm}/b_{c\ell}$.  

Thus, if the bias \bmtol\ is independent of cluster richness, then the lack of
a trend of \mtolmax\ with \nvir\ means we have measured the same asymptotic
\deltamtol\ for all richness bins.  If \bmtol\ is not independent of \nvir,
then by chance the variations in asymptotic \deltamtol\ were canceled by a
corresponding change in the bias.  So we need to determine whether this bias is
constant. 

The bias \bmtol\ should primarily depend on the mass of the halos hosting the
light tracers, and this should be related to the luminosity of those galaxies.
Included in the Tables \ref{tab:m2lngals} and \ref{tab:m2llum} is the mean
$i$-band luminosity of galaxies at $10 h^{-1}$ Mpc separations $\langle
L_{10}^{gal}\rangle$ .  The value on larger scales was not well constrained for
all richness bins.  There is not a strong variation of this mean luminosity
between the richness bins.  The average, over all cluster richness bins, of the
luminosity of galaxies within 10Mpc is \lshift\ = $(1.10 \pm 0.04) \times
10^{10} h^{-2} L_\odot$ ($-20.43 \pm 0.04$ mag).  Note in \S
\ref{sec:ldens_universe} we saw that $L^*_{^{0.25} i} = 1.7 \times 10^{10}
h^{-2} L_\odot$.  Given the small variation in luminosity, and the fact that
the bias varies quite slowly for $L<L_*$ galaxies \citep{Tegmark04}, the bias
\bmtol\ should be roughly constant for each cluster sample.  Thus we will
assume we are measuring the true asymptotic value at large separations, and
average these values from all cluster richness bins.

In section \ref{sec:m2lasym} we calculated this asymptotic value in two ways:
first by taking the value of the integrated \deltamtol\ for the last radial bin
\maxinvertrad\ to get \mtolmax, and the second fitting a simple descriptive
model to get \mtolasym.  Averaging over all cluster luminosity bins gives
\begin{equation}
    \langle M/L \rangle~\textrm{\bmtolinv} = 
        \left\{ \begin{array}{ll}
                 \textrm{\meanmtolrmaxn} & \textrm{within \maxinvertrad} \\
                 \textrm{\meanmtolasymn} & \textrm{asymptotic fit}
                \end{array}
        \right.
\end{equation}
in solar units, where the bias \bmtol\ corresponds to that of \lshift\ $ =1.10
\times 10^{10} h^{-2}$L$_\odot$ galaxies at z=0.25.

In \S \ref{sec:ldens_universe} we calculated that the luminosity density in the
$^{0.25}i$-band is \ldens.  Multiplying the asymptotic M/L above by this, and
dividing by the critical density,  number we get an estimate of $\Omega_m$ that
is independent of $h$:
\begin{equation}
    \Omega_m \textrm{\bmtolinv} = 
        \left\{ \begin{array}{ll}
                 \textrm{\omegarmax} & \textrm{within \maxinvertrad} \\
                 \textrm{\omegaasym} & \textrm{asymptotic fit}
                \end{array}
        \right.
\end{equation}
There is certainly more information than we have used in the \maxinvertrad\
values since points at smaller radius do contain independent information.  So
in principle a more precise measurement could be made, but the error within
\maxinvertrad\ can be considered conservative.  On the other hand, the error
bar on the asymptotic fit is certainly an underestimate, as the fit is not
based on a physical model.  The errors are small simply because all the points
in the curve are used rather than just the last, but in fact the points on 
small scales are not necessarily informative in interpreting the points on 
large scales in absence of a physical model; this error should be thought 
of as a lower limit.

In addition there is a level of systematic error not accounted for here.  
Although the level of systematic error is not precisely known, we expect
it to be $\lesssim 10$\%, mostly due to errors in photometric redshifts
of lensing source galaxies. See \citet{JohnstonLensing07} for a more
complete discussion of systematic errors.

There have been numerous studies measuring mass-to-light ratios of clusters.
As discussed above with regards to the \mtolvir-\mvir\ relation, there are a
wide variety of techniques and conventions in place.   Many of these studies
use the mass-to-light ratio to infer $\Omega_m$ by assuming the value they get
is equal to that of the  universe.  Although this is not always a
well-justified assumption, converting to $\Omega_m$ does remove most of the
dependence on the bandpass, galaxy apertures, and mass aperture (as long as the
mass aperture isn't too small).  Using the inferred $\Omega_m$ may lead to a
more robust comparison between the various results.

The series of papers by Bahcall et al. have consistently estimated $\Omega_m
\sim 0.2$ from this technique, using various mass estimators. For example,
using X--ray clusters \citet{BahcallComerfordM2l02} found $0.17 \pm 0.05$ and
SDSS clusters whose masses were calibrated from velocity dispersions gave $0.19
\pm 0.08$ \citep{BahcallM2l03}.  

Cluster velocity dispersions in the CNOC data have also been used to calculate
$\Omega_m$.  Using stacked velocity and $B$-band light profiles,
\citet{CarlbergM2l97} found $\Omega_m = 0.19 \pm 0.06$.  A recent analysis
using individual masses and $K$-band light found $0.22 \pm 0.02$
\citet{MuzzinCNOCm2l07}.

Using the ``caustic'' method for estimating masses and $K$-band light in the
CAIRNS survey, \citet{RinesCairnsM2l04} inferred $\Omega_m = 0.18 \pm 0.03$.
This method claims to yield more accurate masses by correctly modeling the
infall regions around clusters.  The study of CNOC2 groups by \citet{Parker05}
using weak lensing for group masses and $B$-band light found $\Omega_m = 0.22
\pm 0.06$.  It should be noted that these are rather small groups; as we have
shown in this work the mass-to-light ratio is less than universal within their
virial radii.  Using X--ray data for a set of 2MASS clusters, and 2MASS
$K$-band luminosities, \citet{LinMohrStanford03} find $\Omega_m = 0.19 \pm
0.03$.  

Although there is general agreement when using mass-to-light ratios to infer
$\Omega_m$, these measurements give smaller values for $\Omega_m$ than baryon
fraction measurements, although baryon fraction measurements do have dependence
on $h$.  This is noted, for example, by \citet{LinMohrStanford03} where they
find a higher $\Omega_m = 0.28 \pm 0.03$ using the baryon fraction using their
own data, and suggest their mass-to-light ratios are too low.  

In our analysis we demonstrated that the mass-to-light ratio reaches an
asymptotic value at large radius, which removes one possible error in
determining the global M/L.  The missed light, ICL and otherwise, will
generally push the mass-to-light ratios even lower.  It is possible that the
\bmtol\ is considerably greater than unity, but this is not the theoretical
expectation. The light tracers are $L < L_*$ galaxies which should reside in
under-biased halos \citep{ShethTormen99,SeljakWarren04}.

Note we have assumed a fiducial flat cosmology with $\Omega_m$ = 0.27 when
performing these measurements, and this has not been varied properly in order
to constrain $\Omega_m$ above. Thus the calculated $\Omega_m$ could be slightly
biased, and there could be some additional statistical error not accounted for
here.  Generally, decreasing the assumed $\Omega_m$ increases the inferred
$\Omega_m$ using this technique, through it's affect on the critical density
for lensing. But at these redshifts such effects are secondary, well below our
errors.  In a future paper we will present a full cosmological analysis
including these effects as well as a proper model for BCG displacements from
the halo mass peak as carried out in \citet{JohnstonLensing07}.

These results are a precise test for models of structure and galaxy formation.
The M/L results, coupled with the galaxy color and luminosity distributions as
a function of radius from clusters (sections \ref{sec:lumcolorfunc} and
\ref{sec:radlum}) show in detail how different types of galaxies are
distributed in and around these clusters and how they are clustered relative to
the underlying mass distribution \citep*[see also][]{Hansen07}.  These are the
most basic cross-correlation statistics that can be addressed, and are perhaps
the most precise and powerful statistics that can be measured by a photometric
survey.

It is significant that these measurements were carried out in a purely
photometric data set. Every step of the process uses photometric data only,
from cluster finding, to galaxy cross-correlations, to lensing measurements.
These type of measurements can be carried out in any high-quality survey with
properly chosen bandpasses.  Future surveys such as DES, SNAP, and LSST will
greatly extend these measurements and further challenge our theories of
cosmology and galaxy formation.

\acknowledgements

    
    E.S. thanks the Aspen Center for Physics and the organizers of the
    ``Modeling Galaxy Clustering'' workshop for creating such a productive
    working environment; the majority of this paper was written during his two
    weeks there.

    Thanks to David Hogg for many helpful comments. We are grateful to Roman
    Scoccimarro and Mulin Ding for use of the "Mafalda" computing cluster at
    NYU.  Thanks to Jeremy Tinker and David Weinberg for the model predictions
    and many helpful discussions.

    E.S. was supported by NSF grant AST-0428465. B.K. and T.A.M. gratefully
    acknowledge support from NSF grant AST 044327 and the Michigan Center for
    Theoretical Physics

    The research described in this paper was performed in part at the Jet
    Propulsion Laboratory, California Institute of Technology, under a
    contract with the National Aeronautics and Space Administration.

    Funding for the SDSS and SDSS-II has been provided by the Alfred P. Sloan
    Foundation, the Participating Institutions, the National Science
    Foundation, the U.S. Department of Energy, the National Aeronautics and
    Space Administration, the Japanese Monbukagakusho, the Max Planck Society,
    and the Higher Education Funding Council for England. The SDSS Web Site is
    http://www.sdss.org/.

    The SDSS is managed by the Astrophysical Research Consortium for the
    Participating Institutions. The Participating Institutions are the American
    Museum of Natural History, Astrophysical Institute Potsdam, University of
    Basel, Cambridge University, Case Western Reserve University, University of
    Chicago, Drexel University, Fermilab, the Institute for Advanced Study, the
    Japan Participation Group, Johns Hopkins University, the Joint Institute
    for Nuclear Astrophysics, the Kavli Institute for Particle Astrophysics and
    Cosmology, the Korean Scientist Group, the Chinese Academy of Sciences
    (LAMOST), Los Alamos National Laboratory, the Max-Planck-Institute for
    Astronomy (MPIA), the Max-Planck-Institute for Astrophysics (MPA), New
    Mexico State University, Ohio State University, University of Pittsburgh,
    University of Portsmouth, Princeton University, the United States Naval
    Observatory, and the University of Washington.

    \normalsize

\bibliographystyle{apj}
\bibliography{apj-jour,astroref}

\end{document}